\documentclass[final, 5p, times]{elsarticle}
\usepackage[T1]{fontenc}
\usepackage[latin9]{inputenc}
\usepackage{amsmath}
\usepackage{amssymb}
\usepackage{graphicx}
\usepackage{caption}
\usepackage{esint}
\usepackage[caption=false, font=footnotesize]{subfig}

\usepackage{booktabs}

\journal{NIMA}
\usepackage{siunitx}[=v2]
\usepackage{upgreek}
\usepackage[bookmarks,bookmarksopen,bookmarksdepth=6]{hyperref}
\hypersetup{
    colorlinks=true,
    linkcolor=blue,
    filecolor=magenta,      
    urlcolor=cyan
}

\usepackage{cases}
\usepackage{url}

\biboptions{sort&compress}
\begin{document}

\begin{frontmatter}

\title{Investigation of the Boron removal effect induced by \SI{5.5}{\mega\electronvolt} electrons on highly doped EPI- and Cz-silicon}
\tnotetext[mytitlenote]{Work performed in the frame of the
\href{https://rd50.web.cern.ch/}{CERN-RD50} collaboration.}

\author[mymainaddress]{C.~Liao\corref{mycorrespondingauthor}}
\cortext[mycorrespondingauthor]{Corresponding author}
\ead{chuan.liao@desy.de}

\author[mymainaddress]{E.~Fretwurst}
\author[mymainaddress]{E.~Garutti}
\author[mymainaddress]{J.~Schwandt}
\author[mysecondaryaddress]{L.~Makarenko}

\author[mythirdaddress]{I.~Pintilie\corref{mycorrespondingauthor}}
\ead{ioana@infim.ro}

\author[mythirdaddress]{Lucian~D.~Filip}
\author[myfourthaddress]{A.~Himmerlich}
\author[myfourthaddress]{M.~Moll}
\author[myfourthaddress]{Y.~Gurimskaya}
\author[myfifthaddress]{Z.~Li}

\address[mymainaddress]{Institute for Experimental Physics, University of
Hamburg, Hamburg, Germany}
\address[mysecondaryaddress]{Belorussian State University, Minks, Belarus}
\address[mythirdaddress]{National Institute of Materials Physics, Bucharest,
Romania}
\address[myfourthaddress]{European Organization for Nuclear Research (CERN),
Geneva, Switzerland}
\address[myfifthaddress]{College of physics and optoelectronic engineering,
Ludong University, Yantai, China}

\begin{abstract}
This study focuses on the properties of the B$_\text{i}$O$_\text{i}$ (interstitial Boron~-~interstitial Oxygen) and C$_\text{i}$O$_\text{i}$ (interstitial Carbon~-~interstitial Oxygen) defect complexes by \SI{5.5}{\mega\electronvolt} electrons in low resistivity silicon. Two different types of diodes manufactured on p-type epitaxial and Czochralski silicon with a resistivity of about 10~$\Omega\cdot$cm were irradiated with fluence values between \SI{1e15}{\per\square\centi\meter} and \SI{6e15}{\per\square\centi\meter}. Such diodes cannot be fully depleted and thus the accurate evaluation of defect concentrations and properties (activation energy, capture cross-section, concentration) from Thermally Stimulated Currents (TSC) experiments alone is not possible. In this study we demonstrate that by performing Thermally Stimulated Capacitance (TS-Cap) experiments in similar conditions to TSC measurements and developing theoretical models for simulating both types of B$_\text{i}$O$_\text{i}$ signals generated in TSC and TS-Cap measurements, accurate evaluations can be performed. The changes of the position-dependent electric field, the effective space charge density $N_\text{eff}$ profile as well as the occupation of the B$_\text{i}$O$_\text{i}$ defect during the electric field dependent electron emission, are simulated as a function of temperature. The macroscopic properties (leakage current and $N_\text{eff}$) extracted from current-voltage and capacitance-voltage measurements at \SI{20}{\celsius} are also presented and discussed. 
\end{abstract}


\end{frontmatter}

\section{Introduction}
In order to cope with extraordinary high particle rates up to 200 p--p collisions per bunch crossing in High Luminosity Large Hadron Collider (HL-LHC) experiments, new types of silicon sensors were developed e.g.\ Low Gain Avalanche Detectors (LGADs)~\cite{b1,b2,b3}, and High Voltage CMOS devices (HV-CMOS) for inner tracking detectors~\cite{b4,b5,b6,b7,b8}. Both types of sensors as well as the new pixel and strip devices will be manufactured on boron doped ($p$-type) silicon. The degradation of the performance of these sensors is due to the expected high radiation field. For instance, exposing the LGADs to a particle radiation field leads to the reduction of the internal gain value with increasing fluence. This degradation is caused by a deactivation of the active boron in the highly doped $p$-type gain layer (about \SI{5e16}{\per\cubic\centi\meter}), which leads to a reduction of the space charge and consequently, a lowering of the electric field followed by a decrease in charge multiplication in this layer. 
\par
In general, the deactivation of the boron dopant is a process called boron removal. A possible way to reduce boron removal is a co-implantation of carbon into the gain layer~\cite{b2}. The assumed mechanism behind this effect is the competition between the displacement of substitutional boron (B$_\text{s}$) and substitutional carbon (C$_\text{s}$) by primary silicon interstitials (Si$_\text{I}$) into interstitial positions (B$_\text{i}$) and (C$_\text{i}$), respectively. Both interstitial atoms are mobile at room temperature and can react with different impurities, ending up e.g.\ in the formation of B$_\text{i}$O$_\text{i}$ or C$_\text{i}$O$_\text{i}$ defects~\cite{b30, b9, b10, b33, b27, b36}. Although both defects have donor states in the bandgap of silicon, the B$_\text{i}$O$_\text{i}$ act as a trap for electrons and the C$_\text{i}$O$_\text{i}$ as hole trap. At room temperature (RT) the B$_\text{i}$O$_\text{i}$ is positively charged and its concentration affects the effective space charge density ($N_\text{eff}$) while C$_\text{i}$O$_\text{i}$ is in a neutral charge state with no influence on $N_\text{eff}$. The C$_\text{i}$O$_\text{i}$ defect has an energy level in the lower half of the bandgap of silicon, with an activation energy of 0.36~eV and temperature dependent capture cross sections for holes and electrons \cite{b36}. On the other hand, the B$_\text{i}$O$_\text{i}$ defect is a coulombic center having an energy level in the upper half of the silicon bandgap with activation energy  depending on the electric field, experimentally determined to be between 0.24 and 0.26~eV \cite{b27}, and independent on temperature capture cross sections of $\num{1e-14}$~$\si{cm^{2}}$ for electrons  and $\num{1e-20}$~$\si{cm^{2}}$ for holes \cite{b30,b33}.
\par
The reactions with these defects are still under investigation and are of high relevance for improving the radiation hardness of LGADs. In order to get more information about the introduction of both defects and their interplay as well as a quantitative determination of the boron removal rate, the main goal of this work is to accurately characterize the radiation-induced defect complexes B$_\text{i}$O$_\text{i}$ and C$_\text{i}$O$_\text{i}$, and evaluate the boron removal rate in highly boron-doped silicon diodes with different carbon concentrations.
\par
The investigated n$^+$-p diodes were manufactured on 10~$\Omega \cdot$cm p-type epitaxial silicon (EPI) and Czochralski material (Cz), and exposed to high fluences of 5.5~MeV electrons in the range of (1-6)~$\times~10^{15}$~$\si{cm^{-2}}$. The radiation-induced defect complexes B$_\text{i}$O$_\text{i}$ and C$_\text{i}$O$_\text{i}$ were investigated by means of the Thermally Stimulated Current technique (TSC). One problem in the evaluation of the defect concentrations from the measured TSC spectra arises from the fact that the irradiated low resistivity diodes can only be partially depleted during the temperature scan, due to the limit of the maximal reverse voltage which can be applied. That means, that the depleted volume is beforehand not known for the temperature range of the charge emission of the defects. We show that this problem can be overcome if in addition to TSC experiments Thermally Stimulated Capacitance (TS-Cap) method is employed. This method allows the determination of the depleted volume at any temperature.
\par
The paper is structured as follows. In section 2 the experimental details about the used diodes manufactured on $p$-type epitaxial (EPI)- and Czochralski (Cz)-silicon, the irradiation with \SI{5.5}{\mega\electronvolt} electrons and the methods for the investigation of the macroscopic and microscopic properties of the devices are presented. In section 3 we provide the results of the current-voltage and capacitance-voltage measurements. Next, section 4 is dedicated to TSC and TS-Cap experiments, data simulation and analyses, with a focus on the Boron-Oxygen (B$_\text{i}$O$_\text{i}$) defect complex and its correlation with the boron removal process.

\section{Experimental Details}
\begin{table*}[htbp]
\centering
\caption{Device information}
  \begin{tabular}{@{}lccccc@{}}
   \toprule
   Label & EPI-3 & EPI-7 & EPI-9 & Cz-3 & Cz-7\\
  \midrule
  Initial doping concentration $N_\text{eff, 0}$ ($\si{cm^{-3}}$) &  $\num{1.1e15}$ & $\num{1.1e15}$ & 
   $\num{1.1e15}$ & $\num{1.05e15}$ & $\num{1.05e15}$\\
  Initial resistivity ($\si{\Omega\cdot cm}$) & $\num{\approx10}$ & $\num{\approx10}$ & $\num{\approx10}$
   & $\num{\approx10}$ & $\num{\approx10}$\\
  Electron fluence $\Phi_\text{e}$ ($\si{cm^{-2}}$) & $\num{1e15}$ & $\num{4e15}$ & $\num{6e15}$ & 
  $\num{1e15}$ & $\num{4e15}$\\
  Fluence value $\Phi_\text{eq}$ ($\si{cm^{-2}}$)$^*$ & $\num{3.98e13}$ & $\num{1.59e14}$ & $\num{2.39e14}$ 
  & $\num{3.98e13}$ & $\num{1.59e14}$\\
  Area $A$ ($\si{cm^{2}}$) &  0.0621 & 0.0621 & 0.0621 & 0.029 & 0.029\\
  Thickness d ($\si{\micro\meter}$) & 50 & 50 & 50 & 400 & 400\\
  Carbon concentration [$C_\text{s}$] ($\si{cm^{-3}}$) & $\num{2e15}$ & $\num{2e15}$ & $\num{2e15}$ & $\num{3e16}$ & $\num{3e16}$\\
  Oxygen concentration [$O_\text{I}$] ($\si{cm^{-3}}$) & $\num{1.5e17}$ & $\num{1.5e17}$ & $\num{1.5e17}$ & $\num{1.5e17}$ & $\num{1.5e17}$\\
  \bottomrule
  
 $^*$ 1 MeV neutron equivalent fluence
  
  \end{tabular}

\label{tab:table1}
\end{table*}
All the investigated diodes are produced by the company -- "Transistors" that belongs to Integral~\cite{b11}. Five sets of $n^+$-$p$ silicon diodes with a deep diffusion junction of
\SI{7.2}{\micro\meter} depth were investigated~\cite{b12}. Three of them (EPI-3, EPI-7, EPI-9) are \SI{50}{\micro\meter} thick epitaxial layers grown on a highly
boron-doped Cz substrate of \SI{525}{\micro\meter} thickness and resistivity of
0.006~$\Omega\cdot$cm. Those three sets have the same boron content of
\SI{1.1e15}{\per\cubic\centi\meter} in the epitaxial layer, corresponding to a
resistivity of about 10~$\Omega\cdot$cm. The other two diodes (Cz-3, Cz-7) were
processed on $p$-type Cz silicon with about the same resistivity of
10~$\Omega\cdot$cm and a thickness of about \SI{400}{\micro\meter}. Except for boron the main impurities are oxygen and carbon. According to~\cite{b9} the Cz and EPI diodes have similar oxygen content, of $\sim$\SI{1.5e17}{\per\cubic\centi\meter}, while the carbon content differs, being in the range of 2-\SI{3e16}{\per\cubic\centi\meter} and 1.5-\SI{2e15}{\per\cubic\centi\meter} in Cz and EPI, respectively. All diodes have been manufactured without a guard ring structure~\cite{b12}. The distance between the pad boundary and the chip edge is roughly 100~$\si{\mu m}$ for all diodes.
\par
The irradiation with \SI{5.5}{\mega\electronvolt} electrons was performed at room temperature using the accelerator facility at Minsk. Since the range of 5.5~MeV electrons is much larger than the thickness of the EPI- and the Cz-silicon the distribution of the radiation induced defects is uniform throughout the whole bulk of the material. More detailed information can be found in~\cite{b13}. The achieved fluence values were in the range of (1-6)~$\times~10^{15}$~$\si{cm^{-2}}$. For the calculation of the corresponding \SI{1}{\mega\electronvolt} neutron equivalent values, a hardness factor of 0.0398 was used according to the Non-Ionizing Energy Loss (NIEL) data of I. Jun et al.~\cite{b14}. More detailed information of the investigated diodes is summarized in Table~\ref{tab:table1}.
\par
The macroscopic device performance of the investigated diodes was measured by means of current-voltage ($I$--$V$) and capacitance-voltage ($C$--$V$) characteristics.\ The radiation induced changes in the effective space charge density $N_\text{eff}$ and the full depletion voltage $V_\text{fd}$ were determined from $C$--$V$ measurements at \SI{10}{\kilo\hertz}. The capacitances were measured with a LCR meter in parallel mode.
\par
For the characterization of the radiation induced electrically active defects, the TSC and TS-Cap methods were used~\cite{b15,b16,b17,b18,b19}. The experimental setup consists of a closed cycle helium cryostat Model SRDK-205 (Sumitomo Heavy Industries, Ltd, Japan) equipped with a temperature controller Model 340 (Lake Shore, US) and a Keithly 6517A electrometer with a voltage source. For the TS-Cap a LCR meter 4263B from Hewlett Packard is used. The experimental procedure consists of cooling down the sample under zero bias to low temperatures (typically \SI{10}{\kelvin}) where filling of the defects is performed for 30~s either by forward biasing of the diode (electron and hole injection by injecting 1~mA forward current) or \SI{0}{\volt} filling (only majority carrier (hole) injection). Then, the diode is reverse biased and a temperature scan is then recorded by measuring the diode current (TSC) or capacitance (TS-Cap) during heating up the device with a constant rate of $\beta$ = \SI{0.183}{\kelvin\per\second}~\cite{b18}. It should be mentioned here that the range of the reverse bias was chosen that way that the current density was below the soft breakdown. For example for EPI-3 and Cz-3 $V_\text{bias}$ < 100~V. Isothermal annealing experiments were performed up to 120~min at a temperature of \SI{80}{\celsius} for all irradiated diodes, with the subsequent evaluation of the macroscopic and microscopic properties.

\section{$I$--$V$ and $C$--$V$ characteristics}
\begin{figure*}[htp]
  \centering
    \subfloat[\label{fig1a}]{
      \includegraphics[width=0.5\linewidth]{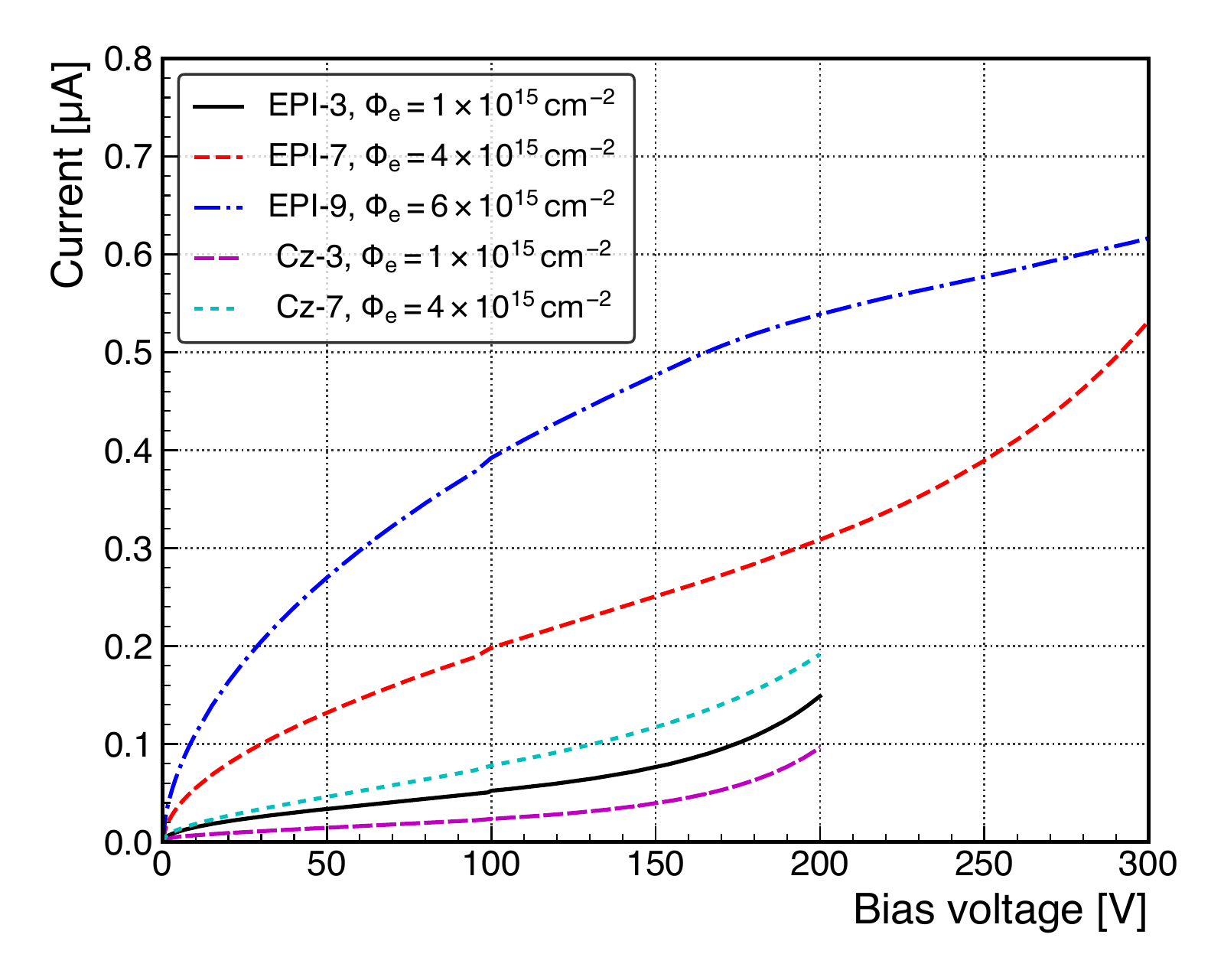}} 
    \subfloat[\label{fig1b}]{
      \includegraphics[width=0.5\linewidth]{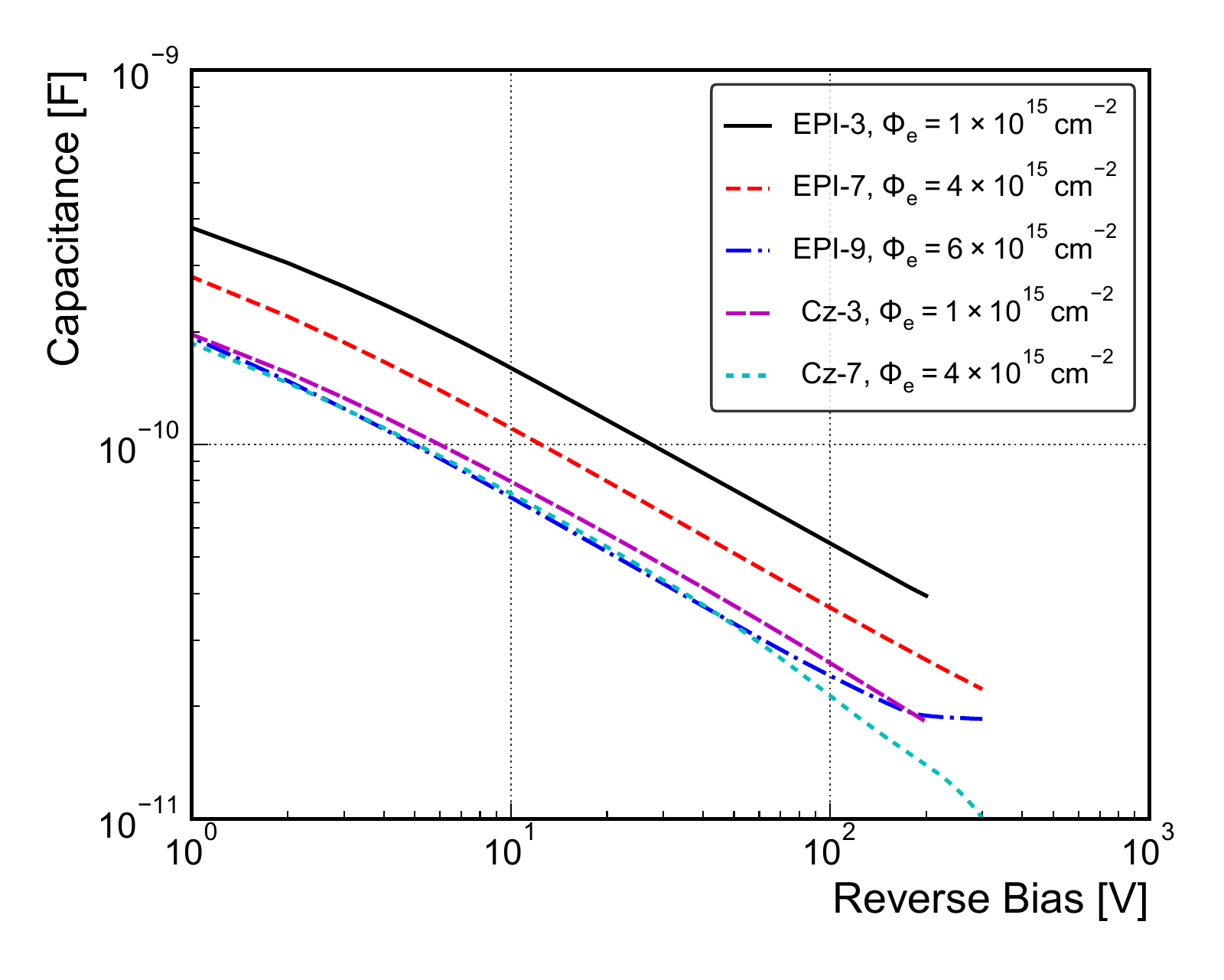}} 
      \\
     \subfloat[\label{fig1c}]{
      \includegraphics[width=0.5\linewidth]{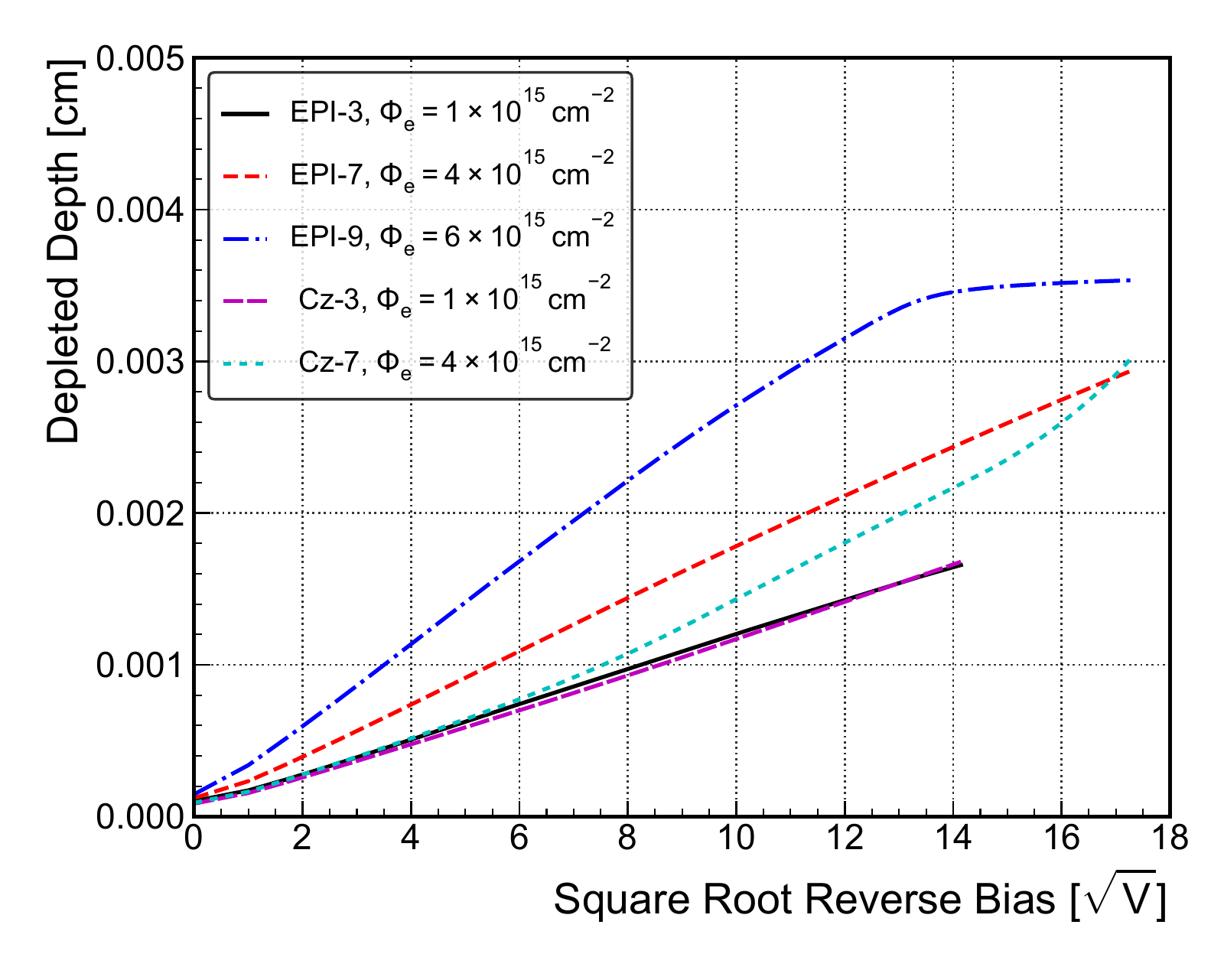}
    }
    \subfloat[\label{fig1d}]{
      \includegraphics[width=0.5\linewidth]{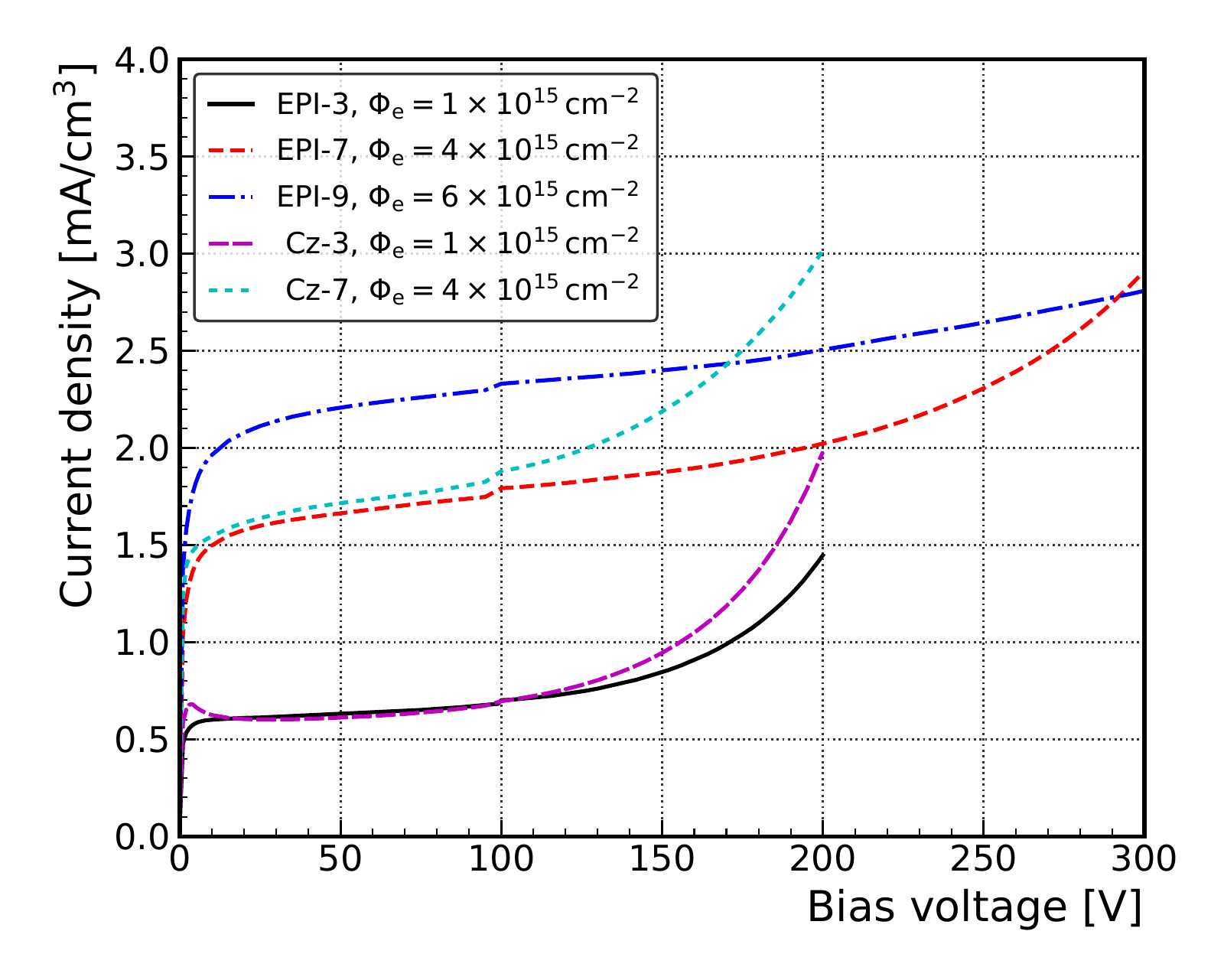}} 
  \caption{(a) Current-voltage characteristics of the 10~$\Omega\cdot$cm
diodes, irradiated with \SI{5.5}{\mega\electronvolt} electrons $\Phi_\text{e}$ =
$1,~4,~6~\times~10^{15}~\si{cm^{-2}}$. Measurements conditions: $T$ 
= \SI{20}{\celsius}, humidity $\leq$ 10\%. (b) $C$--$V$ characteristics and (c) Depleted Depth vs $\sqrt{V}$ of the same 10~$\Omega\cdot$cm diodes presented (a). Measurements conditions: $V_{AC}$ = 0.5~V. Frequency = 10~kHz.
(d) Density of leakage current ($j_d$) versus bias voltage $V$.}
  \label{fig1}
\end{figure*}
\par

In this section, the measured $I$--$V$ and $C$--$V$ characteristics of the irradiated sensors are presented and discussed. As an example, in Fig.~\ref{fig1}(a) the $I$--$V$ curves of all EPI- and Cz-diodes irradiated with different fluences are shown. As it can be seen, for all diodes, except EPI-9 irradiated to \SI{6e15}{\per\square\centi\meter}, a so-called soft breakdown occurs at a certain bias voltage. Such behaviour may have different reasons, e.g.\ the diodes have no guard ring limiting the current to the active pad size and excluding contributions from the outer surface region or edge and/or the high electric field near to the $n^+$-$p$ junction triggers trap assisted Poole-Frenkel or tunnelling effects~\cite{b20,b21}. Nevertheless, determining the depleted depth $w(V)$ from $C$--$V$ characteristics (Fig.~\ref{fig1}(b) and (c)) and assuming that the active area A is given by the $n^+$-pad size, the depleted volume $V_\text{vol}$ = $A \cdot w(V)$ has been calculated and used for estimating the leakage current density $j_d = I/V_\text{vol}$ as a function of the applied bias voltage shown in Fig.~\ref{fig1}(d). One would expect flat curves if edge effects and soft breakdown could be neglected. However, a soft breakdown behaviour is observed in all diodes, except EPI-9. According to Fig.~\ref{fig1}(c), $w$ is plotted as a function of $\sqrt{V}$, in the $\sqrt{V}$ range of (6-10) $\si{V^{0.5}}$ ($V$~between 36~V and 100~V), where the edge effects do not contribute to the rise of the current, $w$ is proportional to $\sqrt{V}$ (except for the diode Cz-7), a typical for the bulk generation current. An average current density $J_d$ was taken in the voltage range from \SI{50}{\volt} to \SI{100}{\volt}, where, a small linear increase of the current is recorded, due to the extension of the electric field in the lateral area of the electrodes in the absence of grounded guard rings. 

\par

\begin{figure}[htb]
\centering
\includegraphics[width=1.0\linewidth]{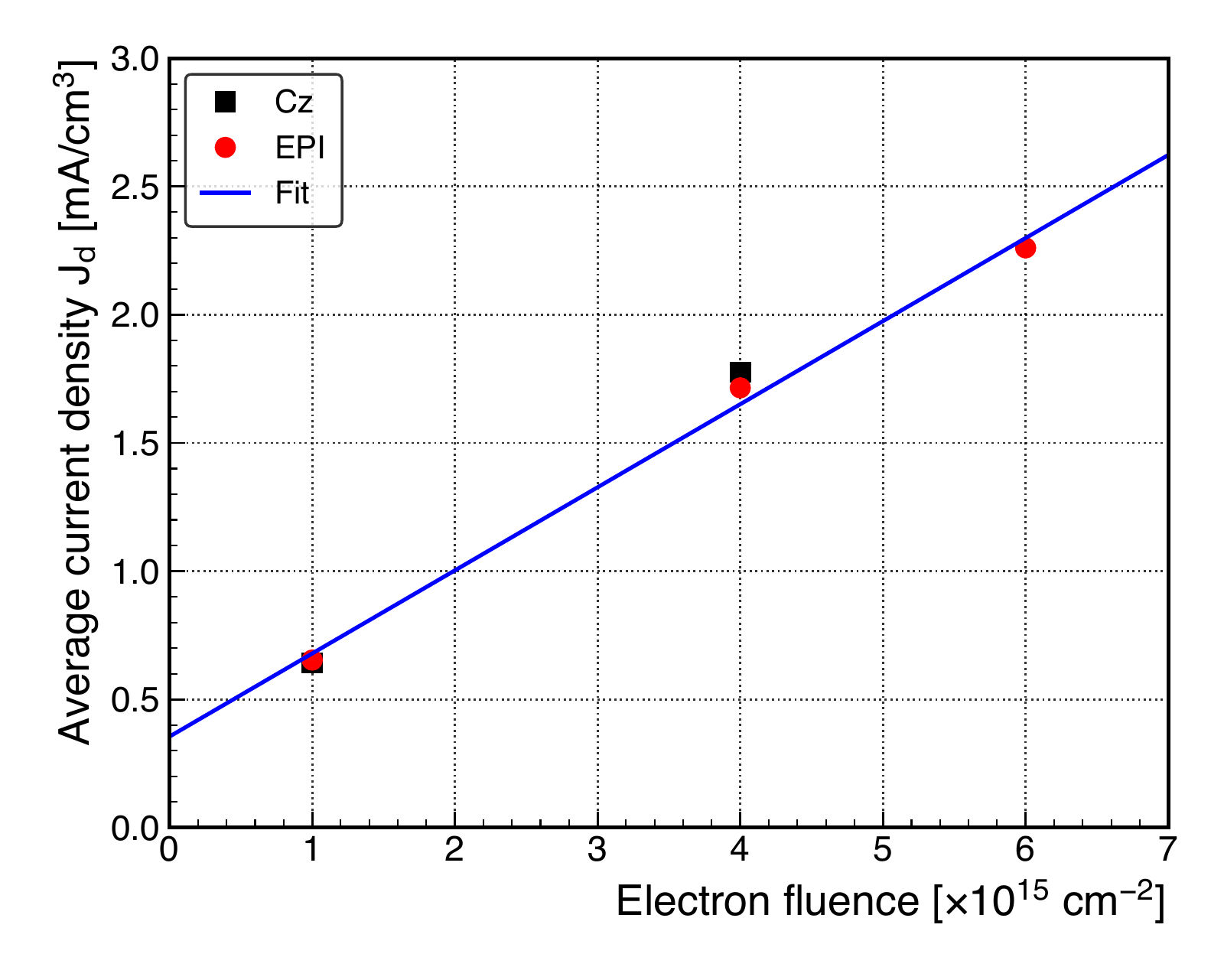}
\caption{Average current density $J_d$ versus. electron fluence (details see text).}
\label{fig2}
\end{figure}

The average values $J_{d}$ of the current densities are plotted as function of the electron fluence $\Phi_\text{e}$ in Fig.~\ref{fig2}, showing an
approximately linear increase. The current related damage
parameter $\alpha$, given by:
\par
\begin{equation}
  \alpha = \frac{\Delta J_d}{\Delta \Phi_\text{e}}
\end{equation}
had been evaluated to $\alpha = (3.2 \pm 0.2) \times 10^{-19}\,\si{\ampere\per\centi\meter}$. Such a small $\alpha$ value was also observed in previous experiments on \SI{5.5}{\mega\electronvolt} electron damage induced in $n$-type silicon~\cite{b22}. Accounting for the hardness factor of 0.0398 the current related damage parameter becomes $\alpha = 0.8 \times 10^{-17}\,\si{\ampere\per\centi\meter}$, being smaller compared to the value of $\alpha = 4\times 10^{-17}\,\si{\ampere\per\centi\meter}$ determined for hadron irradiation and an annealing of 80 min at \SI{60}{\celsius} (see e.g.~\cite{b18,b23,b24}). 
\par

The $C$--$V$ characteristics were measured for 4 different frequencies (\SI{230}{\hertz}, \SI{445}{\hertz}, \SI{1}{\kilo\hertz} and \SI{10}{\kilo\hertz}). A slight frequency dependence is observed and the related explanation can be found in reference~\cite{b25, b26}. The relative deviations measured at 200 V between the values measured at frequencies of \SI{230}{\hertz} and \SI{10}{\kilo\hertz} are below 4\% for all the samples.
\par
\begin{figure}[htb]
\centering
\includegraphics[width=1.0\linewidth]{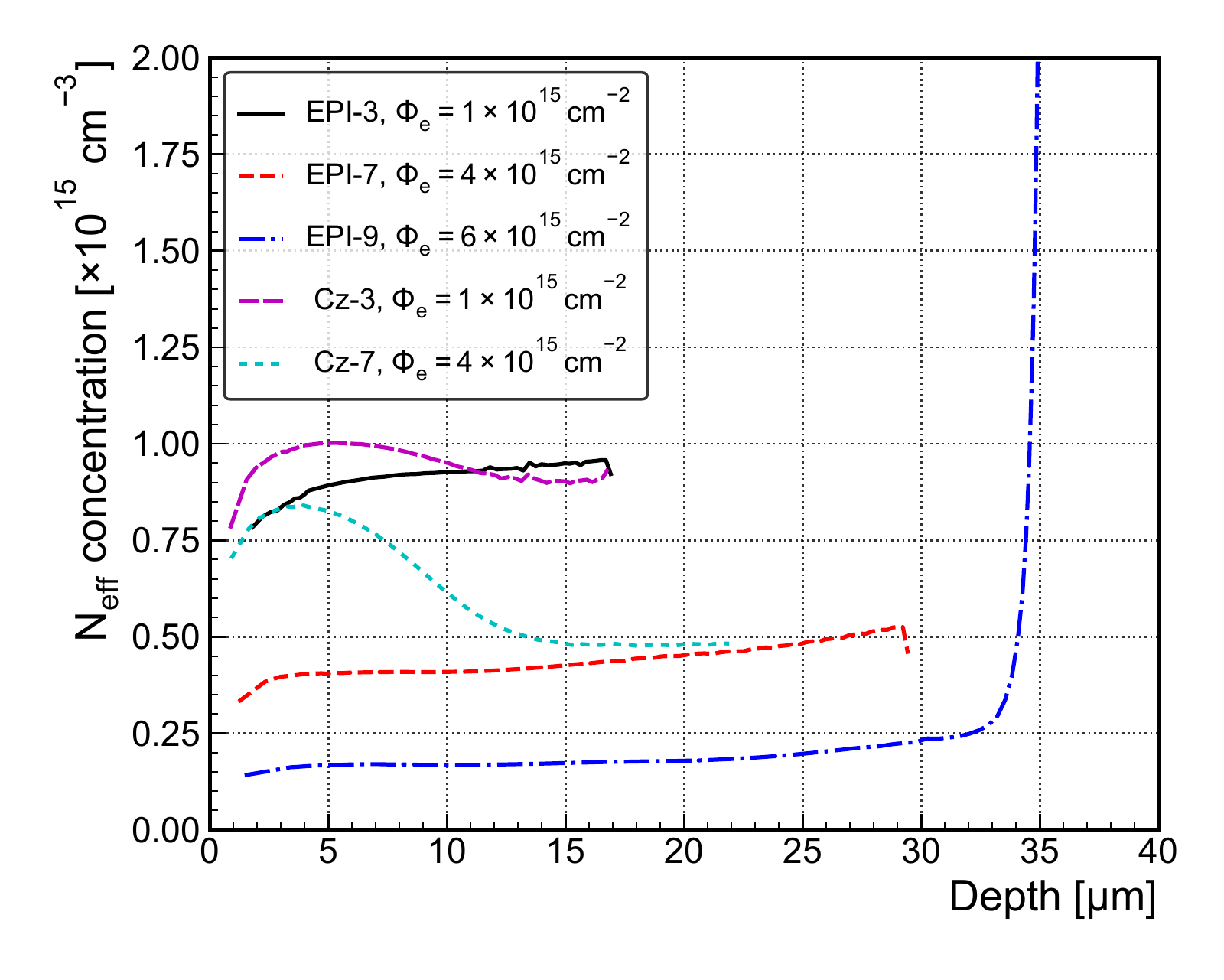}
\caption{$N_\text{eff}$ profile of the diodes irradiated with different fluences. The data were evaluated from $C$--$V$ measurements (Fig.~\ref{fig1}(b)) at room temperature by using Eq.~(\ref{eqn:2}) and Eq.~(\ref{eqn:3}).}
\label{fig3}
\end{figure}

The effective space charge density profile $N_\text{eff}(w(V))$ and the depletion depth
$w(V)$ were extracted from the 10~kHz $C$--$V$ curves (see Fig.~\ref{fig1}(b)) according to Eq.~(\ref{eqn:2}) and Eq.~(\ref{eqn:3}):
\begin{eqnarray}
N_\text{eff} (V) & = & \frac{2}{\epsilon_0 \epsilon_r A^2 q_0\,d(1/C^2)/dV} \label{eqn:2}\\
w (V) & = & \frac{\epsilon_0 \epsilon_r A}{C(V)} \label{eqn:3}
\end{eqnarray}
Where $C$ is the measured capacitance, $\epsilon_0$ is the permittivity of vacuum, $\epsilon_r$ the relative permittivity of silicon (11.9), $q_0$ is the elementary charge, $A$ is the active pad area. Fig.~\ref{fig3} presents the calculated $N_\text{eff}(w(V))$ profiles for the EPI- and Cz-diodes, irradiated with different fluences. 
\par
With increasing fluence, the profiles of $N_\text{eff}$ are shifting to lower values, a fact that is expected mainly due to the deactivation of the initial boron concentration caused by irradiation, the so-called boron removal effect. Of course, some hole traps e.g.\ H(140K) and H(152K) will also affect the space charge density $N_\text{eff}$ but their concentrations are much smaller compared to the concentration of the B$_\text{i}$O$_\text{i}$ defect ([H140K + H152K] $\approx$ $\num{2.5e13}$~$\si{cm^{-3}}$ and [B$_\text{i}$O$_\text{i}$] $\approx$ $\num{4.5e14}$~$\si{cm^{-3}}$ in EPI-9).
\par
\begin{figure}[htb]
\centering
\includegraphics[width=1.0\linewidth]{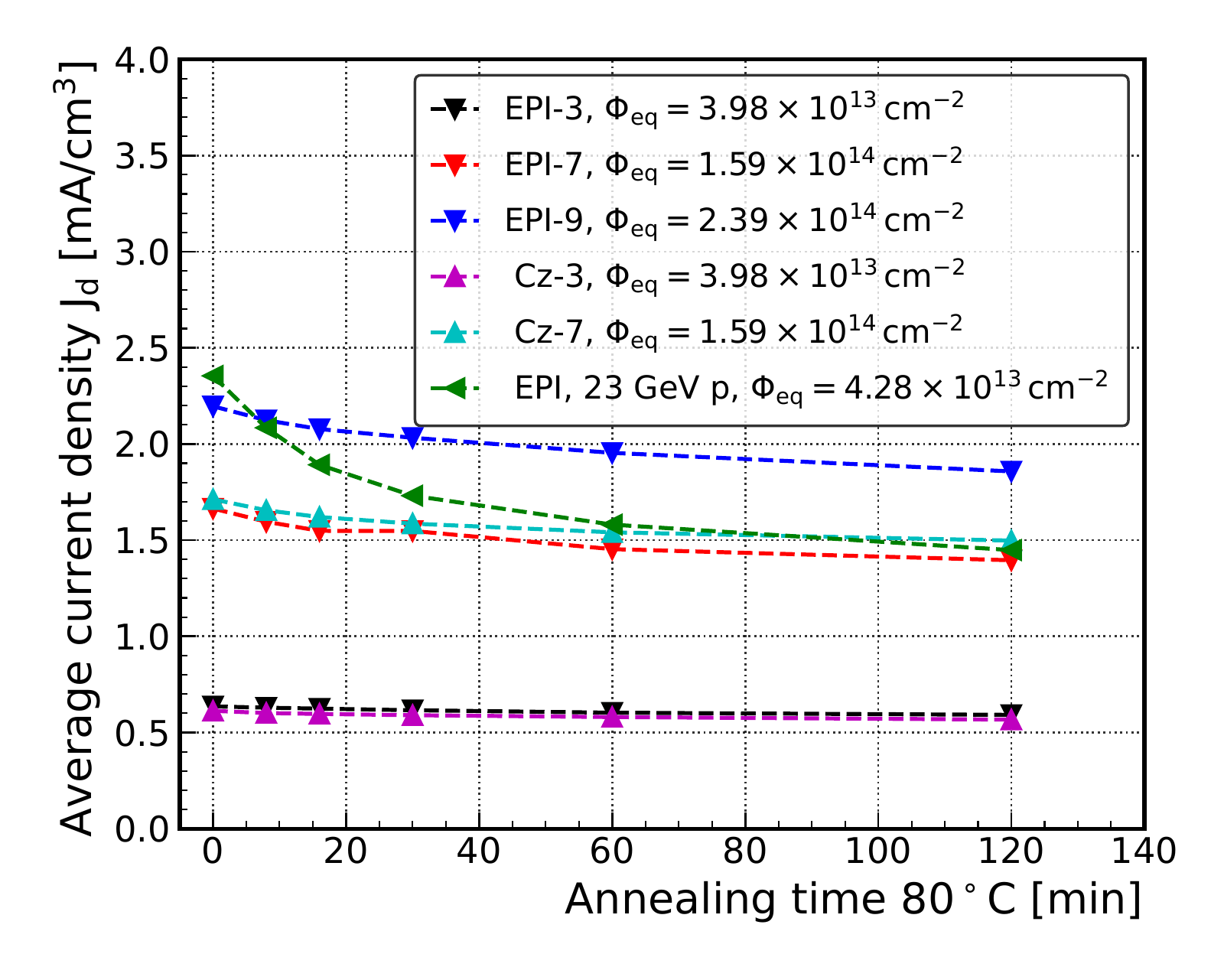}
\caption{$J_d$ versus. annealing times at 80 $^o$C. The data for 23 GeV proton irradiation correspond to a 10~$\si{\Omega\cdot}$cm resistivity EPI-diodes with A = 0.06927~$\si{cm^{2}}$, and have similar $N_\text{eff}$ and $d$ to electron irradiated EPI-diodes.}
\label{fig4}
\end{figure}

The isothermal annealing behaviour of the generation current density $J_{d}$ at \SI{80}{\celsius} is depicted in Fig.~\ref{fig4}. The observed changes with annealing time are much smaller compared to the ones observed for a \SI{23}{\giga\electronvolt} proton irradiated 10~$\si{\Omega\cdot}$cm EPI-diode, which are also included in Fig.~\ref{fig4}~\cite{b27}.
\par
Due to the significant affection by lateral effect especially in Cz diodes, it is deserved to mention the error in the extracted $N_\text{eff}$ and $j_d$. In this work, the lateral effect was estimated by the difference on $j_d$ as shown in Fig.~\ref{fig1}(d), under the assumption that the lateral effect in EPI diodes can be neglected. Thus, for applied
bias voltages of 100~V and 200~V, the error will rise from 0.7\% up to 36\% for Cz-3 and from 5\% to 49\% for Cz-7, respectively. The $V_\text{bias}$ = 100~V corresponds to depleted depths of about \SI{11}{\micro\meter} and \SI{14}{\micro\meter} for the diode Cz-3 and Cz-7, respectively. Only the $J_d$ values from EPI-diodes were used to extract $\alpha$. Thus the error for the $J_d$ was estimated from the bias interval for averaging and resulted in a value of 3\%. This introduces an uncertainty of 5\% in the obtained $\alpha$ value.
\par

\section{Results from TSC and TS-Cap measurements}
\begin{figure*}[htp]
 \centering
  \subfloat[]{
   \includegraphics[width=0.49\linewidth]{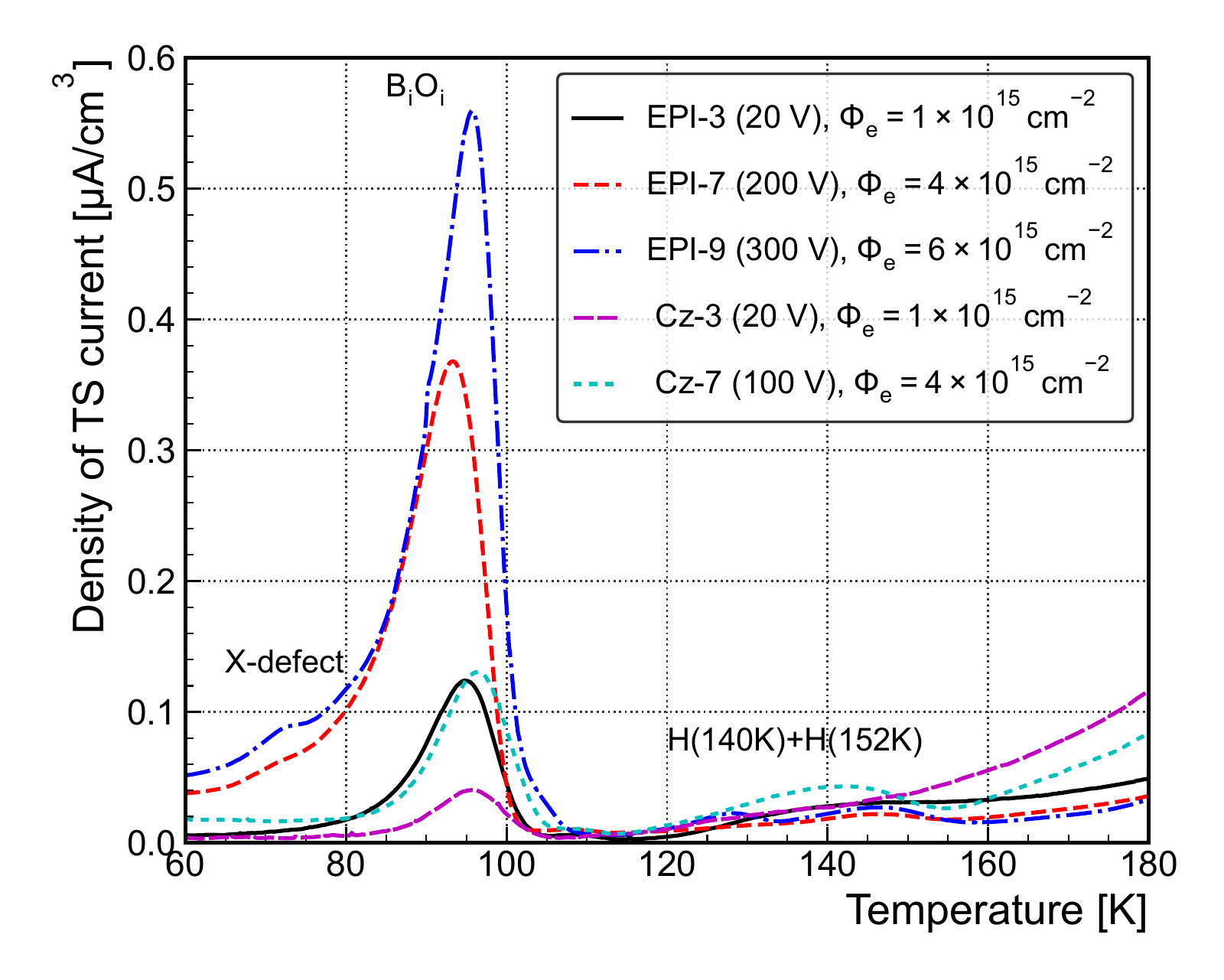}
   }
   \subfloat[]{
   \includegraphics[width=0.49\linewidth]{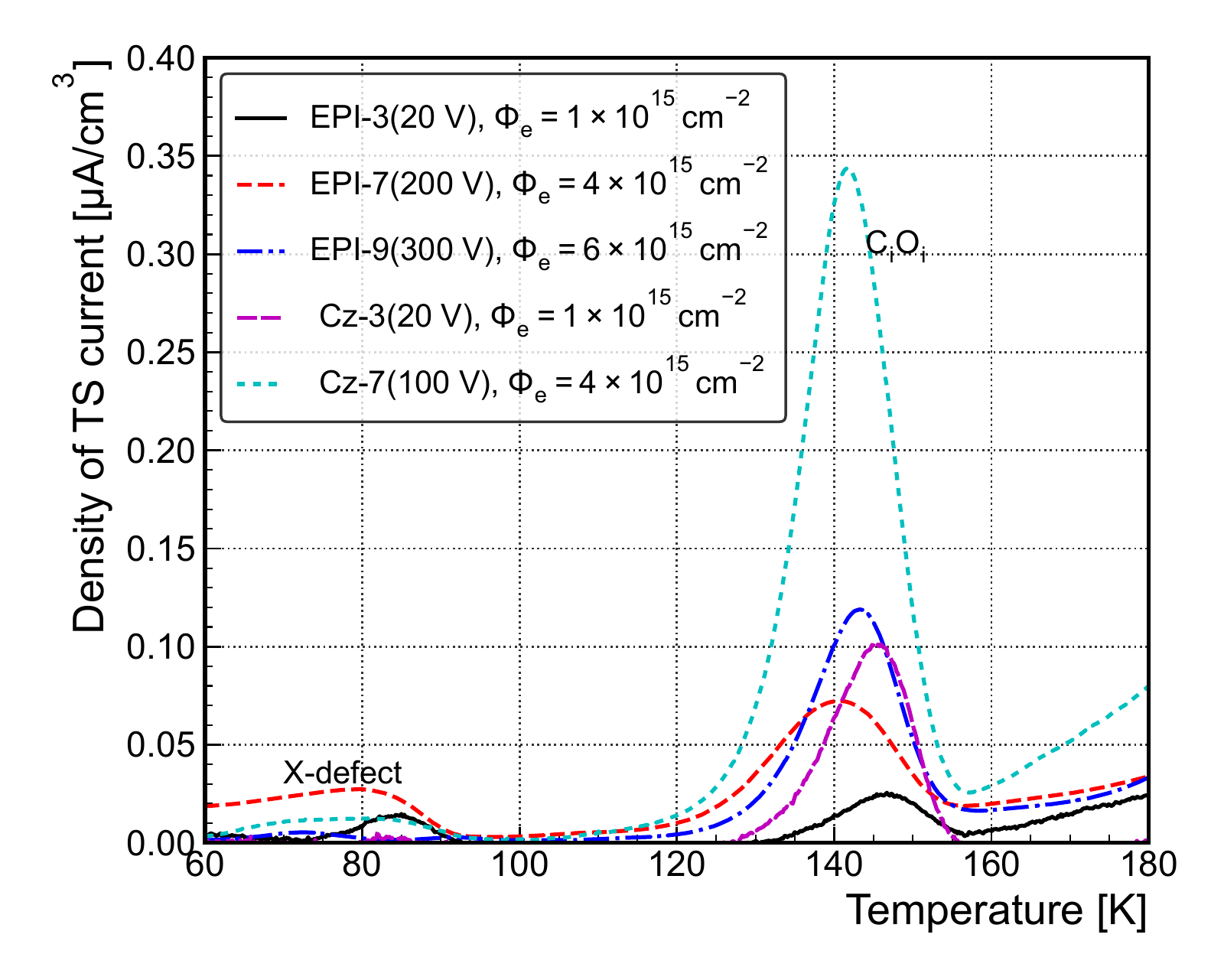}
   } \\
   \subfloat[]{ 
    \includegraphics[width=0.49\linewidth]{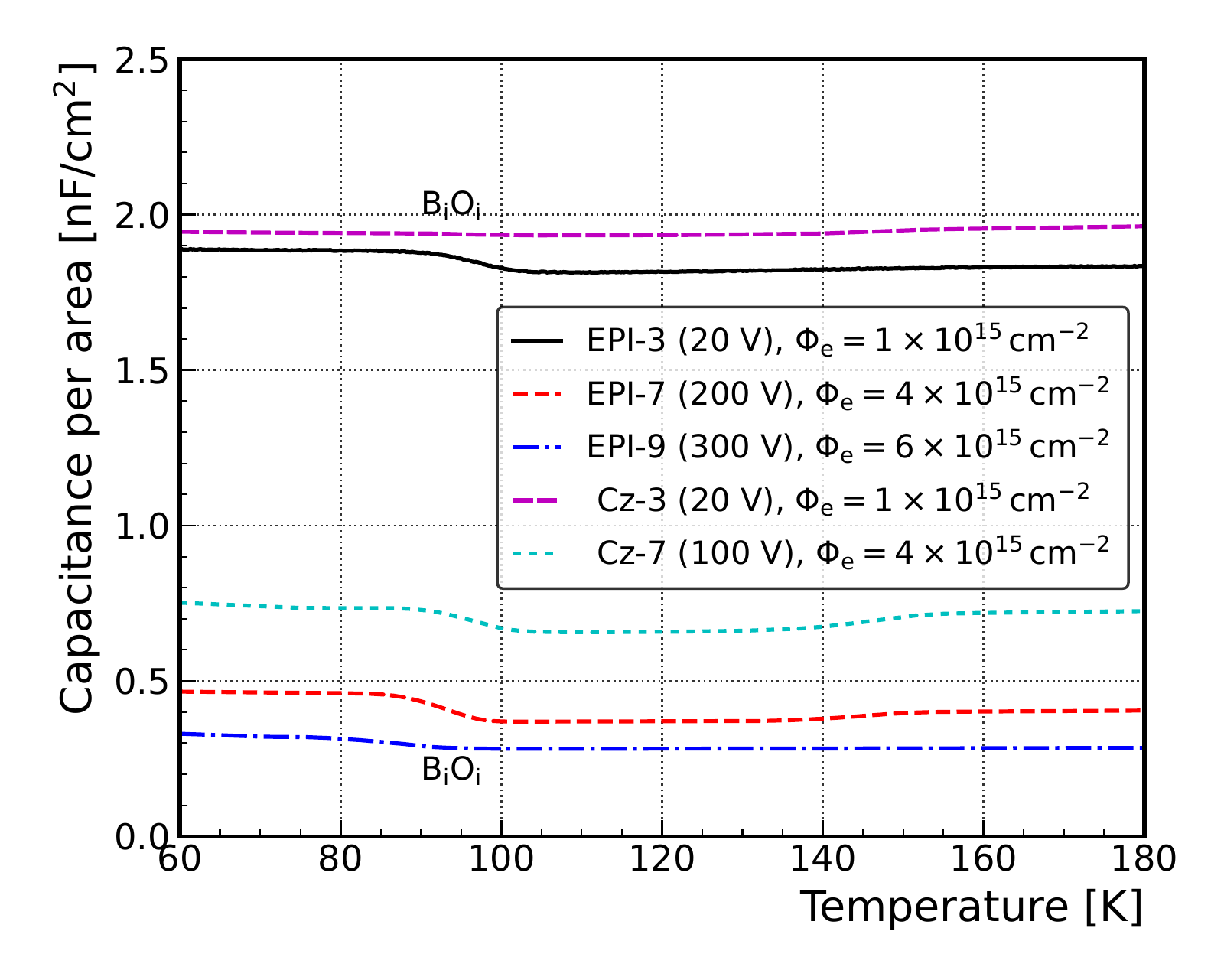}
    }
   \subfloat[]{
      \includegraphics[width=0.49\linewidth]{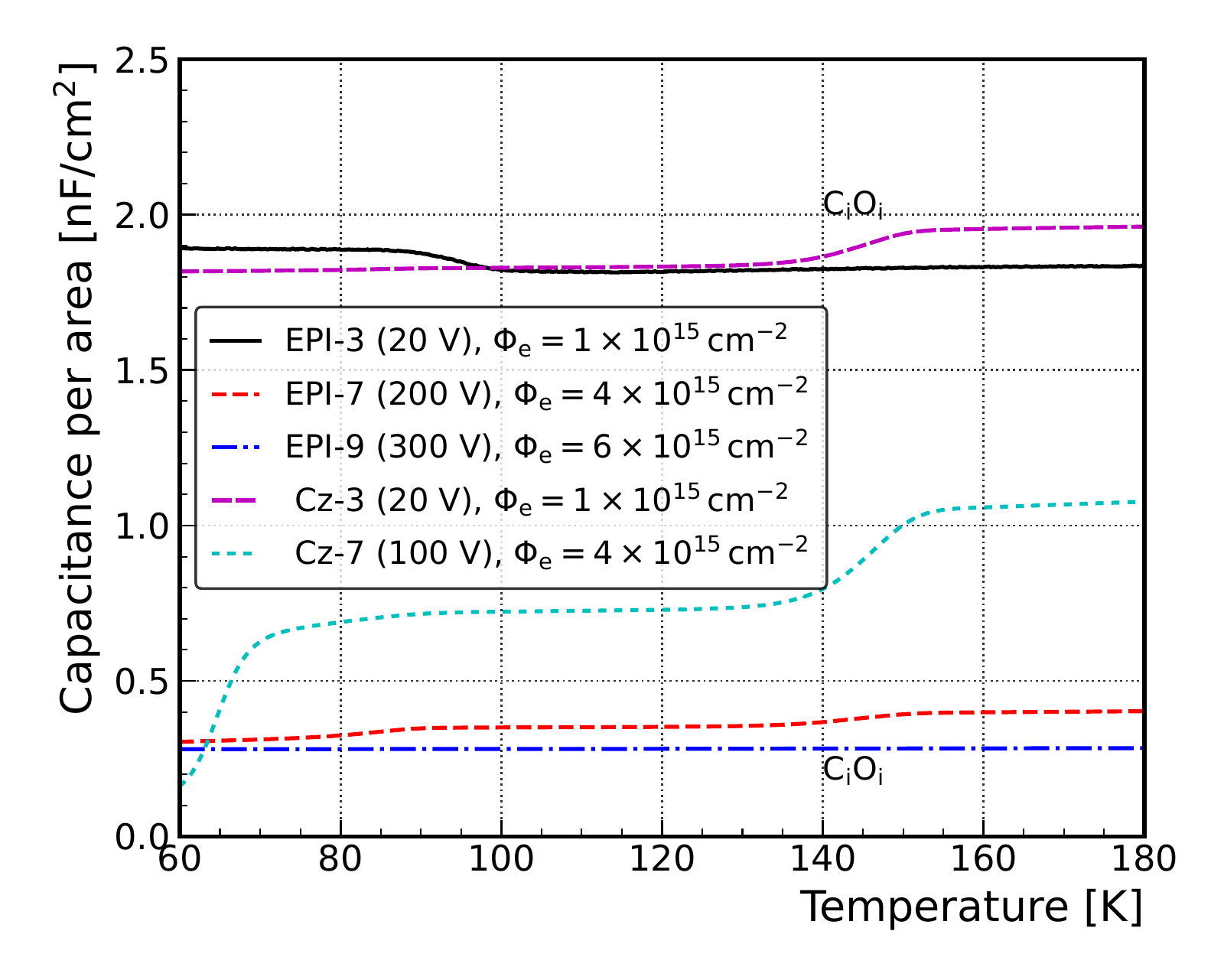}
      } 
\caption{(a) TSC Spectra after trap filling by forward current injection and (b) after filling with majority carriers (holes). Both types of spectra are measured on EPI-(EPI-3, 7, 9) and Cz-diodes (Cz-3, 7) after irradiation with \SI{5.5}{\mega\electronvolt} electrons. The applied bias voltages are indicated in the legends and each diode current is normalized to their individual depleted volume (normalization factor $1/(A \cdot w)$), A~=~active~pad~area, $w(V,T)$~=~depleted~width. (c) and (d) are the TS-Cap measurements corresponding to figures (a) and (b), respectively. The capacitance values are normalized to the pad area of each diode.}
\label{fig5}
\end{figure*}
\begin{figure}[!htb]
\centerline{\includegraphics[width=0.9\linewidth]{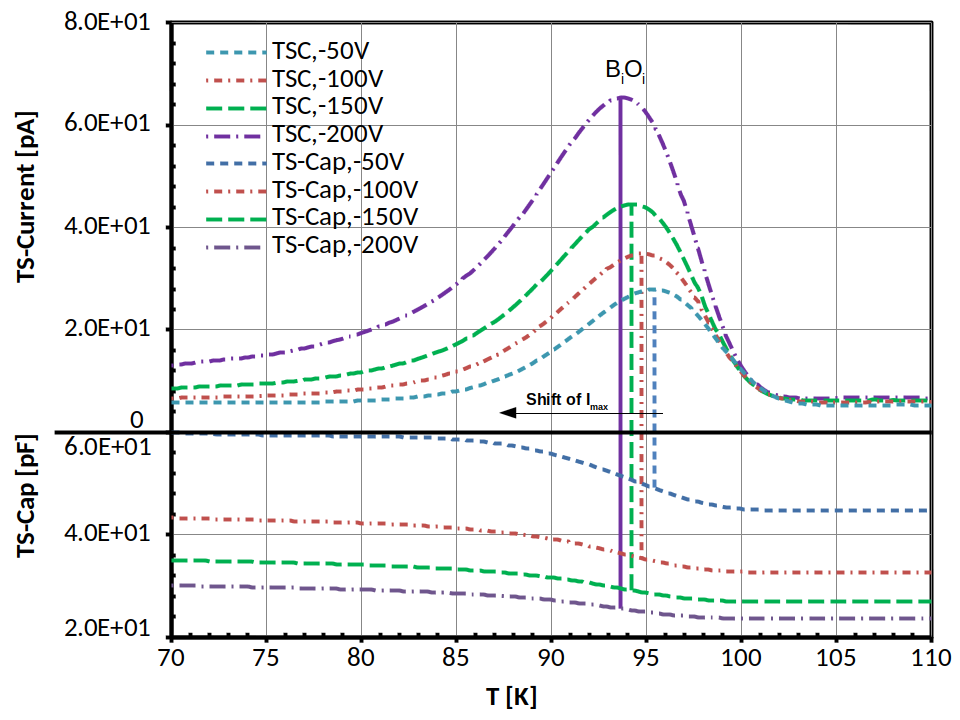}}
\caption{Temperature shift of the B$_\text{i}$O$_\text{i}$ TSC peak in the case of EPI-7 diode ($\Phi_\text{e}$ = $4 \times 10^{15} \si{cm^{-2}}$) for different bias voltages (top) and the corresponding shifts of the TS-Cap curves (bottom). The shifts are indicated by vertical lines between the TSC peak maxima and the turning point of the TS-Cap curves.}
\label{fig6}
\end{figure}
The Thermally Stimulated measurement techniques were used to investigate the defect complexes induced by irradiation with \SI{5.5}{\MeV} electrons, especially the boron-oxygen (B$_\text{i}$O$_\text{i}$) and the carbon-oxygen (C$_\text{i}$O$_\text{i}$) defects in the EPI- and Cz-materials. Figure~\ref{fig5}(a) shows the TSC spectra measured on all diodes (EPI- and Cz-samples) irradiated with different fluences after injecting both electrons and holes (1~mA forward injection) at 10~K. Figure~\ref{fig5}(b) presents the spectra of the same diodes after filling the traps only with holes by cooling the sample to 10~K under 0~V. As can be seen here, the dominant TSC signal occurs at about 150~K and is attributed to the carbon-oxygen (C$_\text{i}$O$_\text{i}$) defect complex. The C$_\text{i}$O$_\text{i}$ signal height in Cz-diodes is much larger compared to the EPI-diodes at the same fluence, due to the higher concentration of carbon in Cz silicon (see Table~\ref{tab:table1}). While Fig.~\ref{fig5}(b) shows only the TSC peaks corresponding to hole traps, Fig.~\ref{fig5}(a) reveals also the ones corresponding to electron traps which can be filled by a forward current injection. As it can be seen in Fig.~\ref{fig5}(a), there is a dominant peak in the temperature range between \SI{90}{\kelvin} and \SI{100}{\kelvin} that is not even traced in the spectra depicted in Fig.~\ref{fig5}(b) corresponding to hole traps only. This dominant peak corresponds to an electron trap, increases with increasing fluence, shows a dependence on the electric field in the sensor, the so-called Poole-Frenkel effect~\cite{b20,b28,b29} as well as a dependence on the impurity content (boron, oxygen, carbon) in the material~\cite{b9,b10,b30,b31} and thus, it is attributed to the B$_\text{i}$O$_\text{i}$ defect complex. Also, theoretical calculations support this identification~\cite{b31}. Because the diodes cannot always be fully depleted during the temperature scan and for a better comparison of the different spectra, the measured currents shown in Fig.~\ref{fig5}(a) and \ref{fig5}(b) had been normalized to the active depleted volume ($V_\text{vol}(V,T)$ = $A \cdot w(V,T)$). The $w(V, T)$ values were extracted from the corresponding TS-Cap measurements.
\par

The TS-Cap data are presented in Fig.~\ref{fig5}(c) and Fig.~\ref{fig5}(d) corresponding to the TSC spectra shown in Fig.~\ref{fig5}(a) and Fig.~\ref{fig5}(b), respectively. For the case of forward current injection the TS-Cap measurements show a drop of the capacitance values in the temperature range of the B$_\text{i}$O$_\text{i}$ emission. This correlates with the change of the B$_\text{i}$O$_\text{i}$ defect charge state, being neutral when occupied with an electron at temperatures before the emission starts and positively charged after the electron is thermally emitted. This leads to a change of the space charge density to a less negative value, corresponding to an increase in the depleted width $w(V,T)$ and consequently to the drop of the capacitance mentioned above. On the other hand, the increase of the capacitance in the range of the C$_\text{i}$O$_\text{i}$ emission (Fig.~\ref{fig5}(d)) is due to the change of the charge state of the C$_\text{i}$O$_\text{i}$ from positive (occupied by holes) to the neutral state after the holes emission. Thus, the space charge density changes from less negative to more negative leading to a decrease in the depleted width $w(V,T)$ and an increase of the capacitance at the given bias voltage. In both cases, the defect concentration can be determined despite the fact that the detector is not fully depleted, as the TS-Cap data can be used to determine the depletion depth at any temperature (see section 4.1.). Further, it is known that the B$_\text{i}$O$_\text{i}$ is a coulombic center~\cite{b30,b31} and thus the electron emission from this defect is governed by the Poole-Frenkel effect, manifesting in a shift of the TSC peak position to lower temperatures with increasing bias voltage. A related shift is then also observed in the TS-Cap curves (see e.g.\ Fig.~\ref{fig6}).
\par
It should be noted that the different values of $V_\text{bias}$ used for different samples were chosen according to the specific characteristics of each diode. Because the aim of the study is to obtain the concentration profiles for defects distributed in the bulk of the diodes, measurements with large $V_\text{bias}$ are preferred in order to scan deep in the bulk of the samples. However, the bias has to be limited to values avoiding the breakdown of the samples. Thus, while the EPI-9 diode withstands a $V_\text{bias}$ =300~V that fully depletes the sample over the entire temperature scan, smaller biases could be applied on the other diodes Thus, the maximum $V_\text{bias}$ that could be safely applied were of 200 V on EPI-7, of 100 V on Cz-7 and of 20 V for Cz-3 and EPI-3. For larger bias values significant increase in the leakage current and dielectric losses at low temperatures were observed.

\par
A quantitative evaluation of defect concentrations from TSC spectra of not fully depleted diodes is only possible if the changes of the depleted depth in the corresponding temperature ranges are known. This issue will be discussed in the following section.
\subsection{Evaluation of concentrations in case of partially depleted sensors}
The TSC method and evaluation of defect properties are described in detail in
numerous publications~\cite{b15,b16,b17,b18,b19}. In our case of not fully depleted devices and the traps homogeneously distributed in the bulk, the current for emission from an isolated electron trap $I^{e}_{TSC} (T)$ with the concentration $n_t(T_0)$, is:
\begin{eqnarray} 
I^{e}_{TSC} (T) & = & q_0 A  n_t(T_0) { \int_0^{w(T)} {x \over {w(T)}}  e_{n}(T, x)  f\left(T, x\right)} dx \\
e_{n} & = & {\sigma_{n} \cdot v_{th,n} \cdot N_{C} \cdot \exp{\left(-{E_a \over
{k_B T}}\right)}}\\
e_{p} & = & {\sigma_{p} \cdot v_{th,p} \cdot N_{V} \cdot \exp{\left(-{E'_{a}
\over {k_B T}}\right)}}
\end{eqnarray}
\begin{eqnarray} 
f(T) & = & {\exp{\left(- {1 \over{\beta}}
\int_{T_0}^{T}{\left(e_n(T')+e_p(T')\right)} \, dT'\right)}}\
\end{eqnarray}
where $T$ is the measured temperature, $w(T)$ the depleted depth at temperature
$T$, $x$ is the coordinate of the depth in the depleted region, $e_n$ and $e_p$ are the emission rates for electrons and holes, respectively, $N_C$ and $N_V$ are the density of states in the conduction band and valence band, respectively. The activation energy for electrons is $E_a$ = $E_C - E_t$ and for holes $E'_a$ = $E_t - E_V$, where $E_t$ is the energy level of the electron traps and $E_{C,V}$ the conduction and valence band edge, respectively. $\sigma_{n,p}$ is the cross section for electrons and holes, $v_{th,n,p}$ is the thermal velocity for electrons and holes. $k_B$ is the Boltzmann constant, $f(T)$ describes the fraction of the defects occupied by electrons at temperature $T$, $\beta$ is the heating rate and $n_t(T_0)$ is the density of defects that are filled with electrons at $T_0$. The $N_C$, $N_V$, $v_{th,n,p}$ values were taken from~\cite{b32} ($N_{C,V} = \num{2.540933e19} \cdot \left({{m^*_{dC,V}} \over {m_0}}\right)^{3/2} \left({{T} \over {300}}\right)^{3/2}$). Eq.~(4) defines the total current which accounts for the conduction and the displacement currents~\cite{b15}. When $f(T)$ and $e_n(T)$ are not position dependent the Eq. (4) can be simplified to:
\begin{eqnarray} 
I^{e}_{TSC} (T) & = & {1 \over 2} \cdot q_0 \cdot A \cdot w(T) \cdot e_{n}(T) \cdot n_t(T_0) \cdot f\left(T\right)  
\end{eqnarray}
\par
In the investigated $p$-type diodes, the B$_\text{i}$O$_\text{i}$ defect, on which this study is focusing, is detected in a TSC experiment only if electrons can be injected at low temperature. This is done by forward biasing the diodes at 10~K injecting both electrons and holes. According to~\cite{b33} the capture cross section for holes of the B$_\text{i}$O$_\text{i}$ defect is neglectable compared with the capture cross section of electrons and thus, $n_{t,0}$ is equal to the defect concentration $N_t$, and $e_p$ can be neglected. Thus, the B$_\text{i}$O$_\text{i}$ defect concentration can be determined by integrating the TSC corresponding signal after filling with forward bias given by Eq. (8) and considering the depleted volume:
\par
\begin{equation}
   N_t = \frac{2}{\beta q_0} \cdot \int_{T_s}^{T_e} \frac{I^{e}_{TSC}(T)}
   {A \cdot w(T)}\,dT =  \frac{2}{\beta q_0} \cdot \int_{T_s}^{T_e} j_{tsc}(T)\,dT
\end{equation}
where $j_{tsc}$ is the thermally stimulated current density, $T_s$ and $T_e$ are the temperature of the start and the end of the electron emission of the defect, respectively. It should be mentioned here that Eq.~(9) is only valid if the defect concentration and the emission rate are position independent. For the investigated irradiated diodes, three different situations have to be considered when evaluating the B$_\text{i}$O$_\text{i}$ concentration: 
\par
(i) At the lowest fluence of $\num{1e15}$~$\si{cm^{-2}}$ the diodes EPI-3 and Cz-3 are partially depleted before and after emission of the defect for all the applied bias voltages. As it can be observed in Fig.~\ref{fig5}(c) the capacitance stays nearly constant, i.e.\ also the depletion depth $w(T)$ is constant in the temperature range of interest. Therefore, Eq.~(9) can be simplified to:
\par
\begin{equation} 
N_t = \frac{2}{\beta q_0} \cdot \int_{T_s}^{T_e} \frac{I^{e}_{TSC}(T)}
   {A \cdot w}\,dT = \frac{2 \cdot Q}{q_0  A  w}
\end{equation}
Where $w$ can be extracted from TS-Cap data as an average value in the range $T_s$ to $T_e$. 
\par
(ii) The sensor is partially depleted before emission and fully depleted after emission. This holds for the device EPI-9, which was irradiated to $\Phi_\text{e}$~=~$\num{6e15}$~$\si{cm^{-2}}$. In this case, the concentration can be evaluated from the TSC spectrum only if $w(T)$ is extracted from TS-Cap measurements.
\par
(iii) Similar to case (i), the sensors are partially depleted before and after emission, but $C(T)$ or $w(T)$ shows visible changes in the temperature range where the electron emission from the defect takes place(see Fig.~\ref{fig5}(c) and \ref{fig5}(d) for the diodes EPI-7 and Cz-7). In this case, the corresponding defect concentration can be directly extracted from the TS-Cap measurement as described in the following.
\par
For high defect concentration where the change in the occupancy of the defects due to the thermal emission of captured electrons or holes leads to measurable variations of the capacitance with increasing temperature, the TS-Cap method can be used to extract the defect concentration. For the B$_\text{i}$O$_\text{i}$ defect the TS-Cap can be described, in the 1-D approach, by the following equations:
\par
\begin{equation}
C(T) =  \frac{ \epsilon_0  \epsilon_r  A}{w(T)}  
\end{equation} 
with
\par
\begin{equation}
w^2 (T) = \frac{2  \epsilon_0  \epsilon_r  (V+V_{bi})}{q_0  
\cdot \lvert N'_\text{eff} (T) \rvert}
\end{equation} 
where
\par
\begin{equation} 
N'_\text{eff}(T) =  N_0 - N_t \cdot (1-f(T))
\end{equation}
Here $C(T)$ is the capacitance of the device at temperature $T$ and for a given
bias voltage $V$, $V_{bi}$ is the build-in voltage, which is negligible compared to the applied bias voltage $V$. The term $N_0$ in Eq.~(13) denotes the absolute $N_\text{eff}$ value before the start of the electron emission of B$_\text{i}$O$_\text{i}$, i.e when all defect centers are neutral and their contribution to the effective space charge concentration is 0. The second term in Eq. (13) accounts for the donor character of the B$_\text{i}$O$_\text{i}$ defect, becoming positively charged after thermal emission of captured electrons and thus leading to a progressive reduction of $N_0$ with increasing the temperature until the electron emission from the defect ends. Assuming no other defects with similar emission rates are present, [B$_\text{i}$O$_\text{i}$] is given by: 
\begin{eqnarray} 
[\text{B}_\text{i}\text{O}_\text{i}] & = & {{2  \epsilon_0  \epsilon_r  V} \over {q_0 }} \left({{1} \over {w^2(T_s)}} - {{1} \over {w^2(T_e)}}\right)
\end{eqnarray}
Here $w(T)$ is extracted from Eq.~(11) and $T_s$ and $T_e$ are the temperatures before and after the electron emission from B$_\text{i}$O$_\text{i}$, respectively.
\par
\begin{figure}[!htb]
  \centering
    \subfloat[]{
     \includegraphics[width=0.9\linewidth]{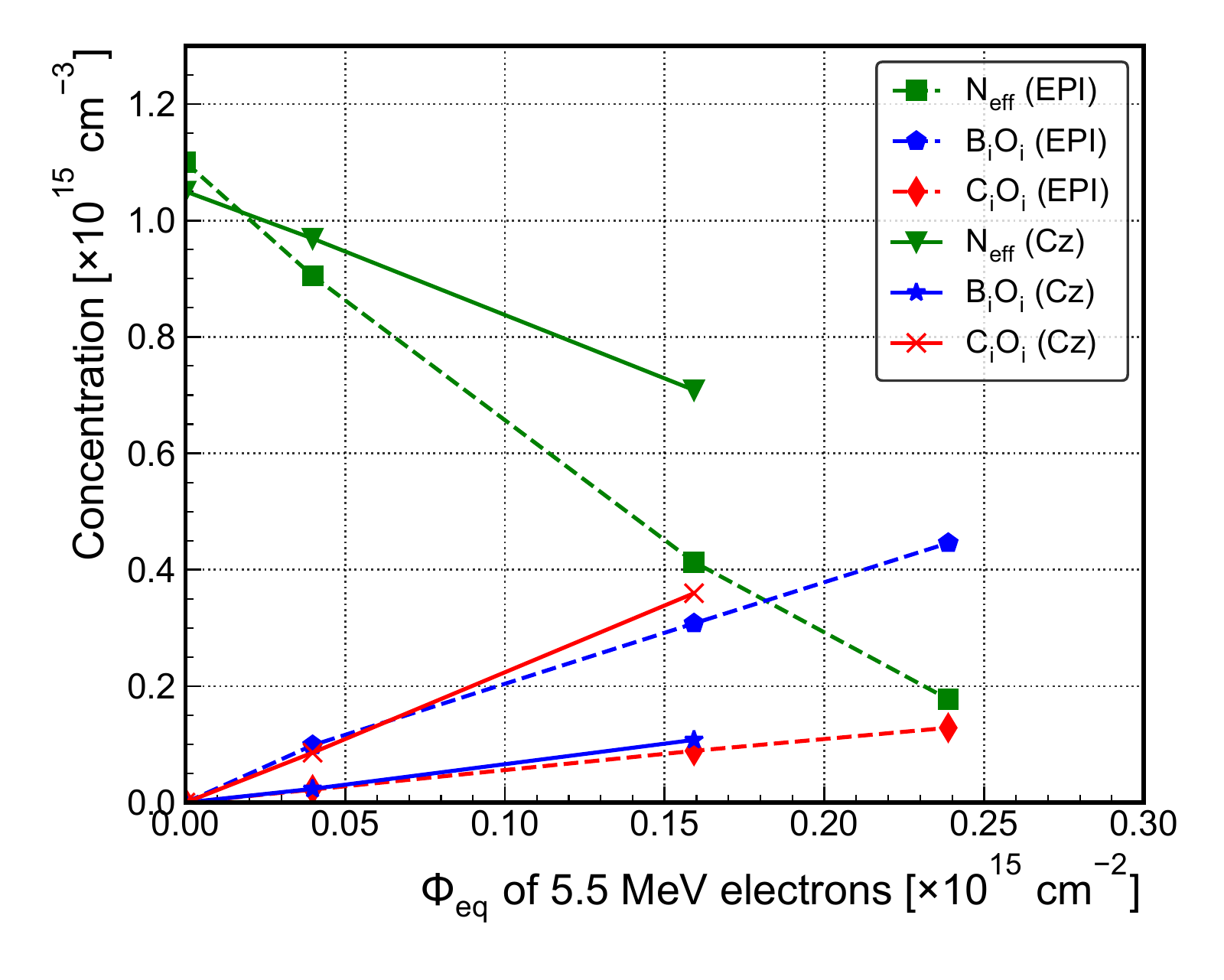}     
    }\\
    \subfloat[]{
     \includegraphics[width=0.9\linewidth]{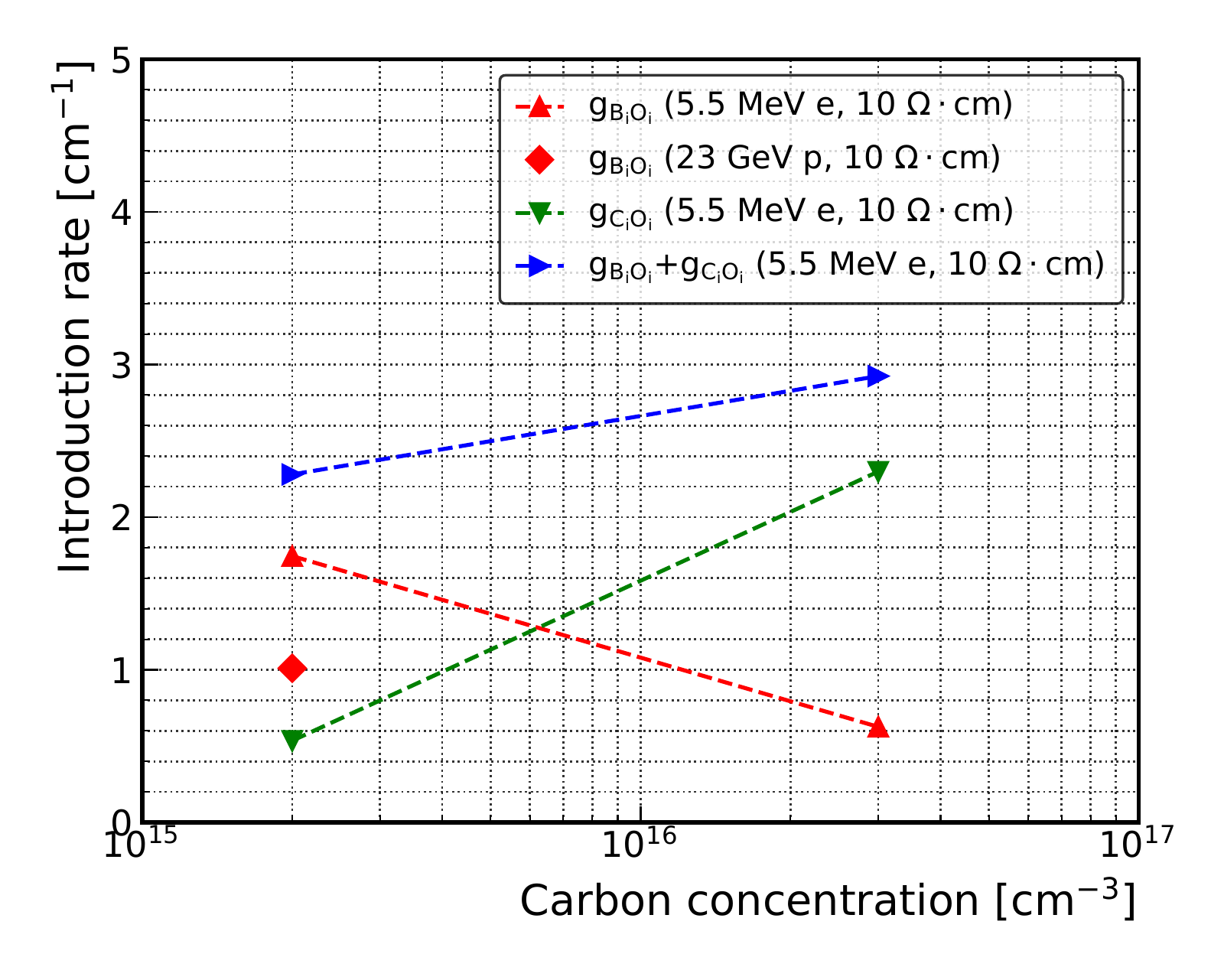}
    }
\caption{(a) Dependence of $N_\text{eff}$, B$_\text{i}$O$_\text{i}$ and C$_\text{i}$O$_\text{i}$ defect concentration on the $\Phi_\text{eq}$ of \SI{5.5}{\mega\electronvolt} electrons for EPI- and Cz- diodes. The $N_\text{eff}$ values were extracted from Fig.~\ref{fig3} in the bias range of 1-100~V and 1-20~V for EPI and Cz diodes, respectively. (b) Variation of $g_{B_i O_i}$ and $g_{C_i O_i}$ as a function of the Carbon content for EPI- and Cz- diodes. Included is  the $g_{B_i O_i}$ value after irradiating a 10 $\Omega\cdot$cm EPI diode with 23 GeV protons at $\Phi_\text{eq}$ = $\num{4.3e13}~\si{cm^{-3}}$.}
\label{fig7}
\end{figure}

In Fig.~\ref{fig7}(a) the B$_\text{i}$O$_\text{i}$ and the C$_\text{i}$O$_\text{i}$ concentrations extracted from the TSC and TS-Cap measurements as a function of $\si{\Phi_\text{eq}}$ are plotted for EPI- and Cz-materials. They were extracted via Eq.~(9) in the temperature range 80-105~K for [B$_\text{i}$O$_\text{i}$] and 120-155~K for [C$_\text{i}$O$_\text{i}$]. Included are also the N$_\text{eff}$ values for both materials as extracted from $C$--$V$ measurements performed at room temperature. The N$_\text{eff}$ values were extracted from Fig.~\ref{fig3} and averaged in the bias range of 1-100 V and 1-20 V for EPI and Cz diodes, respectively. It can be seen from Fig.~\ref{fig5}(c) and Fig.~\ref{fig5}(d) that after carrier emission from B$_\text{i}$O$_\text{i}$ and C$_\text{i}$O$_\text{i}$ the capacitance remains almost constant, and presumably it is the same as at RT. Therefore, using the $N_\text{eff}$ data from RT is appropriate and the introduced errors are related to $N_\text{eff}$ averaging only. The concentrations of B$_\text{i}$O$_\text{i}$ and C$_\text{i}$O$_\text{i}$ defects that can introduce positive space charge in the diodes are lower than the negative charge provided by the Boron-dopant. Therefore, $N_\text{eff}$ remains negative in the entire scanned temperature range.
\par
Assuming the boron removal rate $R$ is given by $R$ = $\lvert (\Delta N_\text{eff})/ (\Delta \Phi_\text{eq}) \rvert$, the values of 2.18~$\si{cm^{-1}}$ and 3.7~$\si{cm^{-1}}$ are obtained for Cz and EPI diodes, respectively. These values were extracted from the slope (absolute value) of the linear fits presented in Fig.~\ref{fig7}(a). The difference of 41\% between the Cz and the EPI rates is attributed to the different amounts of carbon content in both materials as given in Table~\ref{tab:table1}. For the EPI-diodes the change of $N_\text{eff}$ with fluence is roughly a factor 2 larger compared with the increase of the B$_\text{i}$O$_\text{i}$ concentration. This can be explained by the boron removal process, i.e.\ the negatively charged substitutional boron B$^{-}_\text{s}$ is transformed into a positively charged B$_\text{i}$O$^{+}_\text{i}$ defect (B$^{-}_\text{s}$ $\rightarrow$ B$_\text{i}$O$^{+}_\text{i}$). For the Cz-material this cannot be stated due to the strong non-uniform profile of the space charge density (see Fig.~\ref{fig3}). The introduction rates $g_{B_i O_i}$ = [$B_i O_i$]/$\Phi_\text{eq}$ and $g_{C_i O_i}$ =[C$_\text{i}$O$_\text{i}$]/$\Phi_\text{eq}$ were extracted from the linear increase with fluence, and are plotted in Fig.~\ref{fig7}(b) as a function of the carbon content in the EPI- and Cz-diodes. It is obvious that the generation rate of the B$_\text{i}$O$_\text{i}$ is much lower for the material with the higher carbon content. On the other hand, the increase of the C$_\text{i}$O$_\text{i}$ generation rate with increasing carbon content is an indication for the beneficial effect of the carbon impurity in reducing the creation of B$_\text{i}$O$_\text{i}$. This dependence on the carbon concentration has led to the approach of carbon co-implantation into the gain layer of LGADs in order to improve their radiation hardness~\cite{b2}.

\par
Included in Fig.~\ref{fig7}(b) is also the introduction rate of B$_\text{i}$O$_\text{i}$ for an EPI-diode with the same $N_\text{eff, 0}$ and irradiated with the same $\Phi_\text{eq}$ of \SI{23}{\giga\electronvolt} protons as the irradiation with \SI{5.5}{\mega\electronvolt} electrons. As it can be seen, the generation rate of B$_\text{i}$O$_\text{i}$ defect after \SI{5.5}{\mega\electronvolt} electron irradiation is about a factor 1.6 larger than the value determined after irradiation with \SI{23}{\giga\electronvolt}
protons.
\par
In principle, both TSC and TS-Cap are performed with $V_\text{bias}$ where the lateral effect is not significant. However, the obtained concentrations strongly depend on the integration ranges of the TSC spectra or the selection of $T_s$ and $T_e$. Thus, in this work, the error of the extracted B$_\text{i}$O$_\text{i}$ concentrations is given by varying $T_s$ from 75~K to 80~K. The obtained errors for EPI-3, 7, 9 are 5\%, 8\% and 9\%, for Cz-3 and 7 are 5\% and 6\%, respectively. The slightly increasing errors are caused by the overlapping peak at the low temperature tail possibly related to the X-defect. The estimated errors of the $N_\text{eff}$ value shown in Fig.~\ref{fig7}(a) are due to the selected interval of averaging the data (see Fig.~\ref{fig3}). They are about 3\% for all EPI diodes and 5\% for Cz-3. For Cz-7 the estimated error is 20\% due to the non-uniform profile. 
\par
\subsection{Simulation of TSC and TS-Cap data for the B$_\text{i}$O$_\text{i}$ defect}
Compared to the TSC and DLTS methods the TS-Cap technique is rarely used to get
information about radiation induced defects. However, when high concentrations of defects are involved, the method delivers important information on the changes in the depletion depth during a temperature scan from 10~K up to room temperature, which can be used, via developing simulation models, to determine the defect type (capturing electrons or holes) and trapping parameters (activation energy, capture cross section of the emitted charge) as well as its concentration. In our simulations the following assumptions are made:
\par
\begin{itemize}
  \item Lateral effects are neglected.
  \item The device is partially depleted in the temperature range of
interest.
  \item The series resistance of the non-depleted part of the device can be
neglected.
\end{itemize}
\par

\begin{figure}[!htb]
\centerline{\includegraphics[width=0.9\linewidth]{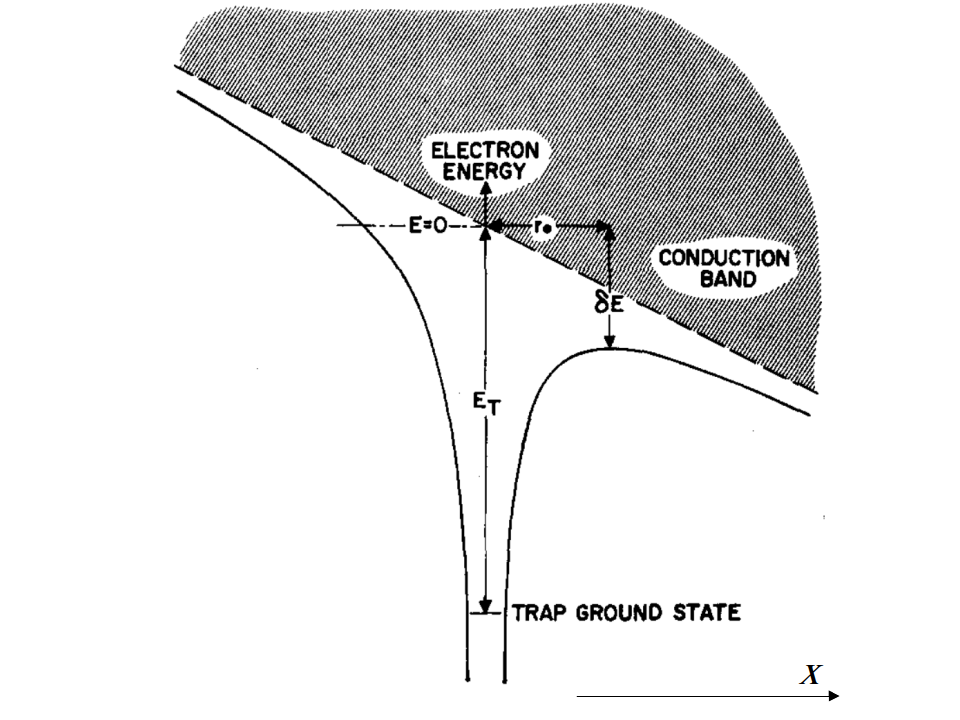}}
\caption{Energy of an electron bound to a positive point charge in the presence of a uniform applied field with the direction $x$ along the bulk \cite{b29}.}
\label{fig8}
\end{figure}

Because the B$_\text{i}$O$_\text{i}$ is a coulombic trap center, the emission rate $e_n$ is not
anymore a constant quantity with respect to the applied bias voltage, but field dependent. By accounting for the 3-D Poole-Frenkel effect, the emission rate can be expressed by~\cite{b20,b28,b29}:
\par
\begin{equation} 
e^{pf}_{n} (T) = e_{n,0} (T) \cdot \left[\left(\frac{1}{\gamma^2}\right)\left(e^{\gamma}\left(\gamma-1\right)+1\right)+ \frac{1}{2}\right]
\end{equation}
where
\par
\begin{equation}
\gamma = \sqrt{q_0
\lvert \vec{E} \rvert \over {\pi \varepsilon_0 \varepsilon_r}} \cdot {q_0 \over
{k_B T}}
\end{equation}
\par
and $e_{n,0}$ denotes the field independent emission rate with the so-called zero
field activation energy $E_a$ = $E_{a,0}$. ${\lvert \vec{E} \rvert}$ is the electric field in the sensor bulk and depends on the position x in the depleted zone. According to the reference~\cite{b29}, the Poole-Frenkel effect is given by the electrostatic energy of an electron which is attracted to a single charged positive ion under the influence of a uniform applied electric field (see Fig.~\ref{fig8}). In the diodes, especially highly doped ones, such an assumption might not be fully valid, since the electric field distribution is not uniform. Thus, in this paper, we introduce a parameter $\xi$ to modify the force between the positively charged ion and the electron. Therefore, the $\gamma$ value is modified to:
\par
\begin{equation}
\gamma = \xi \cdot \sqrt{q_0
\lvert \vec{E} \rvert \over {\pi \varepsilon_0 \varepsilon_r}} \cdot {q_0 \over
{k_B T}}
\end{equation}
In this case the Eq.~(13) has to be revised to:
\par
\begin{equation} 
N^{'}_\text{eff} (T, x) = N_{0} - [\text{B}_\text{i}\text{O}_\text{i}] \cdot (1-f(T,x))
\end{equation}
Furthermore, the electric field distribution $E(T, x)$ in the depleted bulk of the diodes is calculated from the corresponding Poisson equation:
\par
\begin{equation} 
\frac{dE(T, x)}{dx} = \frac{q_0 \cdot N^{'}_\text{eff}(T, x)}{\varepsilon_0 \varepsilon_r}
\end{equation}
The electric field $E$, the occupation fraction f and the $N^{'}_\text{eff}$ are temperature and position dependent. For coulombic centers, the emission rate $e^{pf}_{n}$ has to be used for calculating the occupation fraction defined in Eq. (7).
\par
Considering the involved set of equations, an analytical solution for simulating the TSC and TS-Cap experimental data will be extremely complicated. Therefore, the finite element method is used for simulating the experimental data. The details are presented in the Appendix.

\begin{table*}[htbp]
\centering
\caption{Parameters of simulation. $E(T,x)$ represents the position and temperature dependent electric field and $<E(T)>$ is the average electric field in the diodes}
  \begin{tabular}{@{}lcccccc@{}}
   \toprule
   Methods & TS-Cap ($E(T, x)$) & TSC ($E(T, x)$) & TS-Cap ($E(T, x)$) & TSC ($E(T, x)$) & TS-Cap ($<E(T)>$) & TSC ($<E(T)>$)  \\
   Irradiation & Proton & Proton & Electron & Electron &  Electron & Electron  \\
  \midrule
  $N_\text{0}$ (at 80~K) ($\si{cm^{-3}}$) &  $\num{1.1e14}$ & $\num{1.1e14}$ & 
   $\num{5.1e14}$ & $\num{5.1e14}$ & $\num{5.1e14}$ & $\num{5.1e14}$ \\
   
   [B$_\text{i}$O$_\text{i}$] ($\si{cm^{-3}}$)* &  $\num{3.5e13}$ & $\num{3.3e13}$ &  $\num{2.3e14}$ & $\num{1.6e14}$ & $\num{2.3e14}$ & $\num{1.6e14}$ \\
   $E_{a0}$ ($\si{eV}$)* & 0.265 & 0.273 & 0.258 & 0.258 & 0.284 & 0.284 \\
   $\sigma_n$ ($\si{cm^2}$) & $\num{1.0e-14}$ & $\num{1.0e-14}$ & $\num{1.0e-14}$ & $\num{1.0e-14}$ & $\num{1.0e-14}$ & $\num{1.0e-14}$\\
   Area $A$ ($\si{cm^{2}}$) &  0.06927 & 0.06927 & 0.0621 & 0.0.0621 & 0.0.0621 & 0.0.0621\\
   $\xi$* & 0.85 & 0.85 & 0.5 & 0.5 & 1 & 1\\
  \bottomrule
  \end{tabular}
\begin{flushleft}
Adjusted parameters are indicated by *
\end{flushleft}

\label{tab:table2}
\end{table*} 

\begin{figure*}[htp]
 \centering
  \subfloat[]{
   \includegraphics[width=0.49\linewidth]{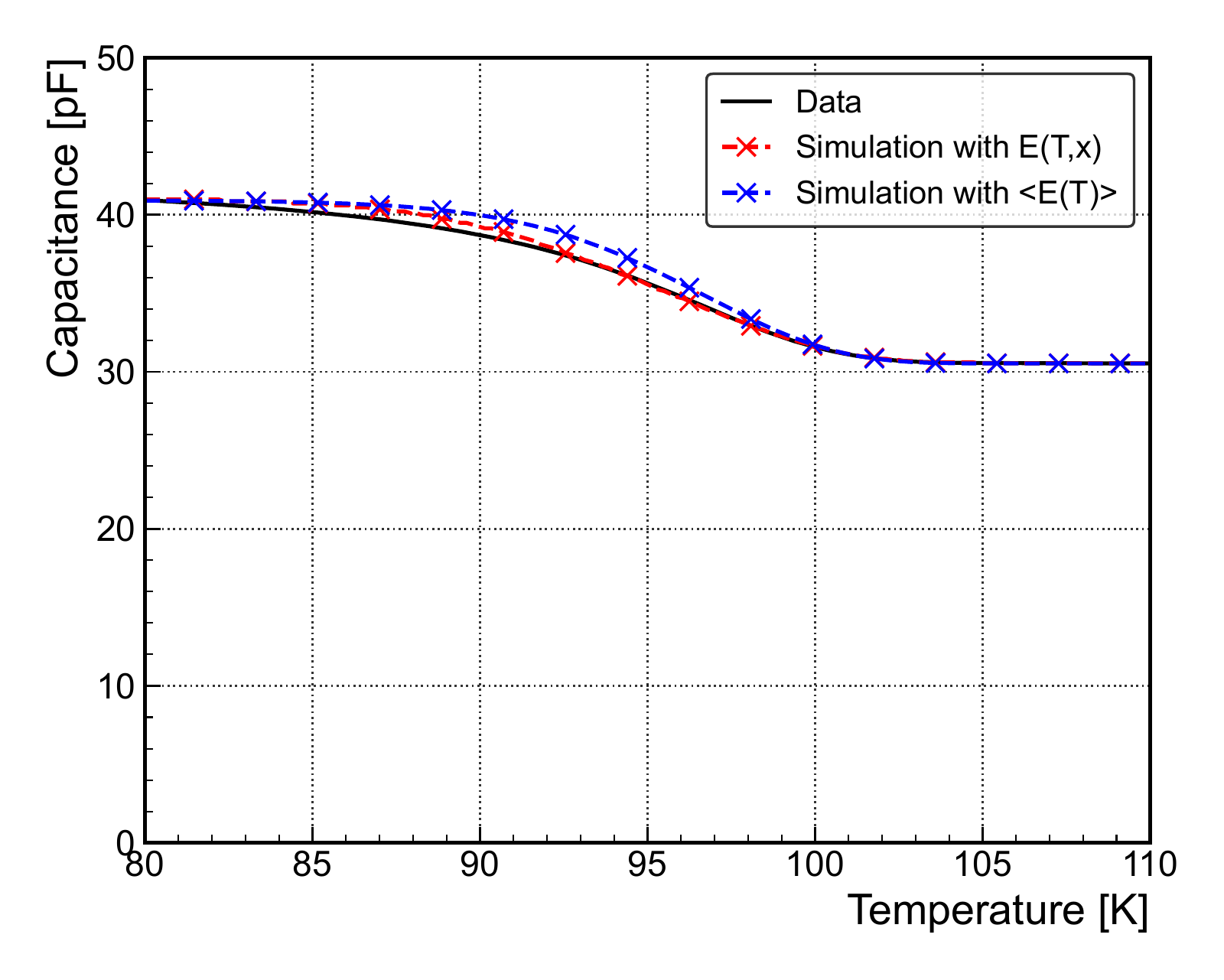}
   }
   \subfloat[]{
   \includegraphics[width=0.49\linewidth]{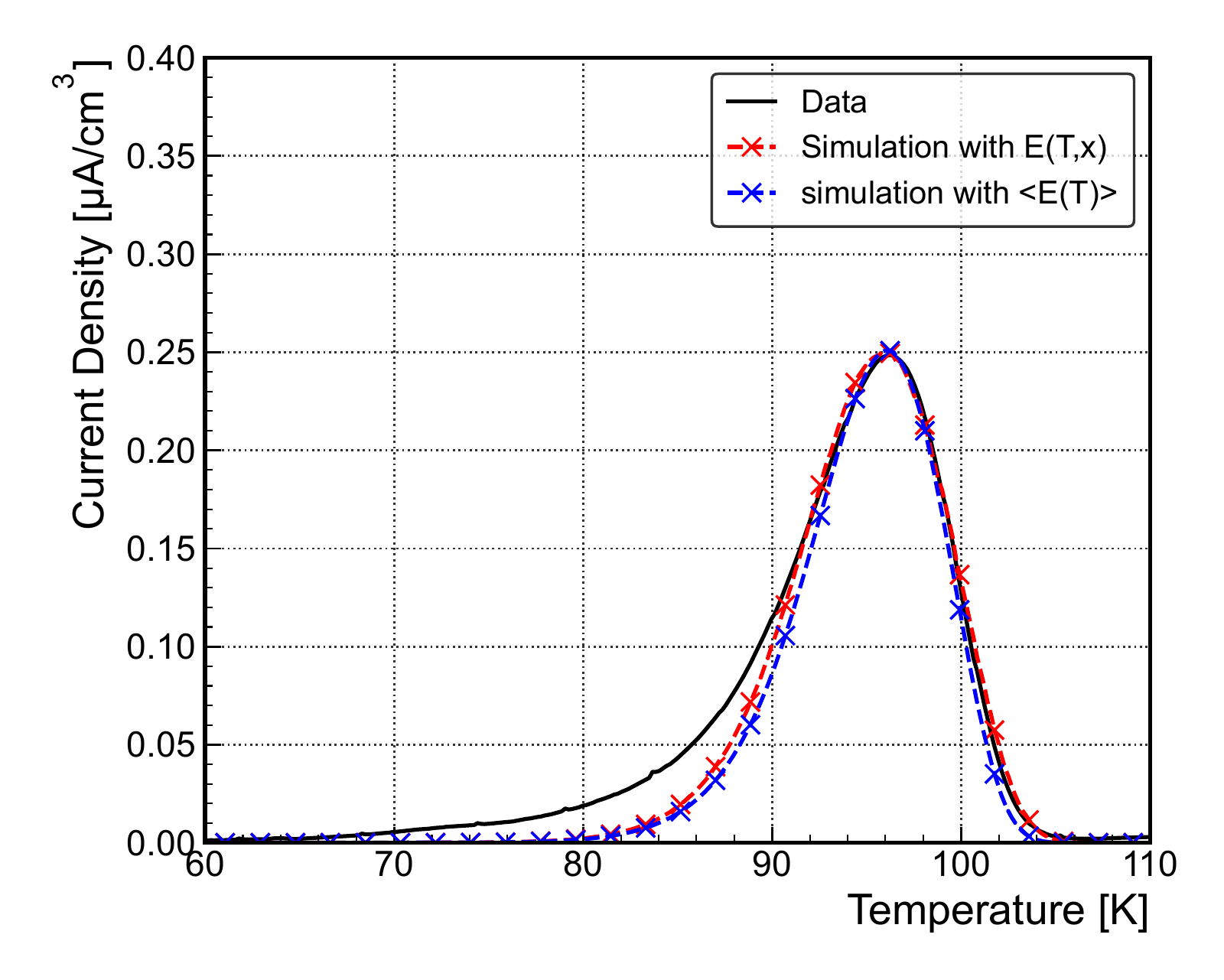}
   } \\
   \subfloat[]{ 
    \includegraphics[width=0.49\linewidth]{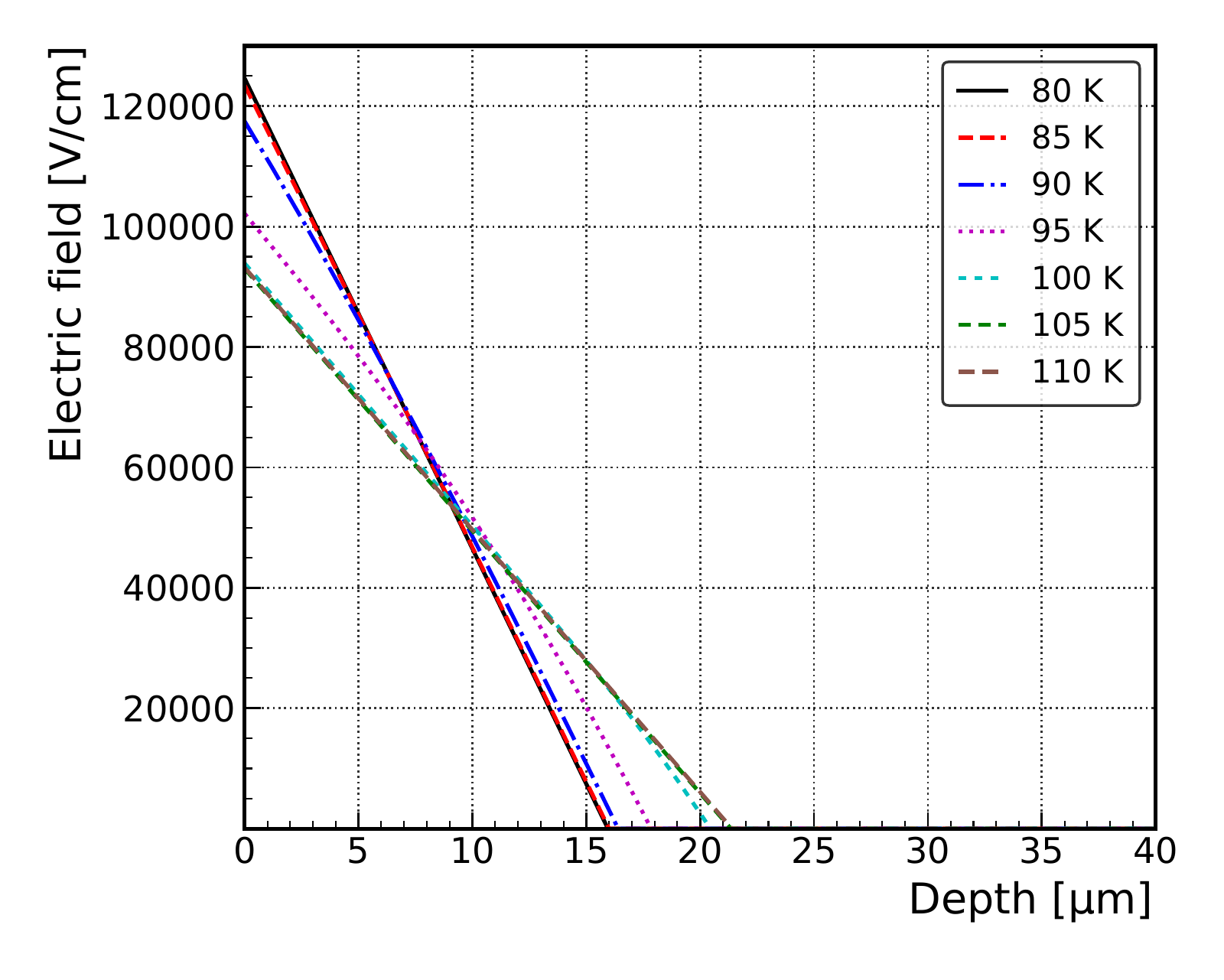}
    }
   \subfloat[]{
      \includegraphics[width=0.49\linewidth]{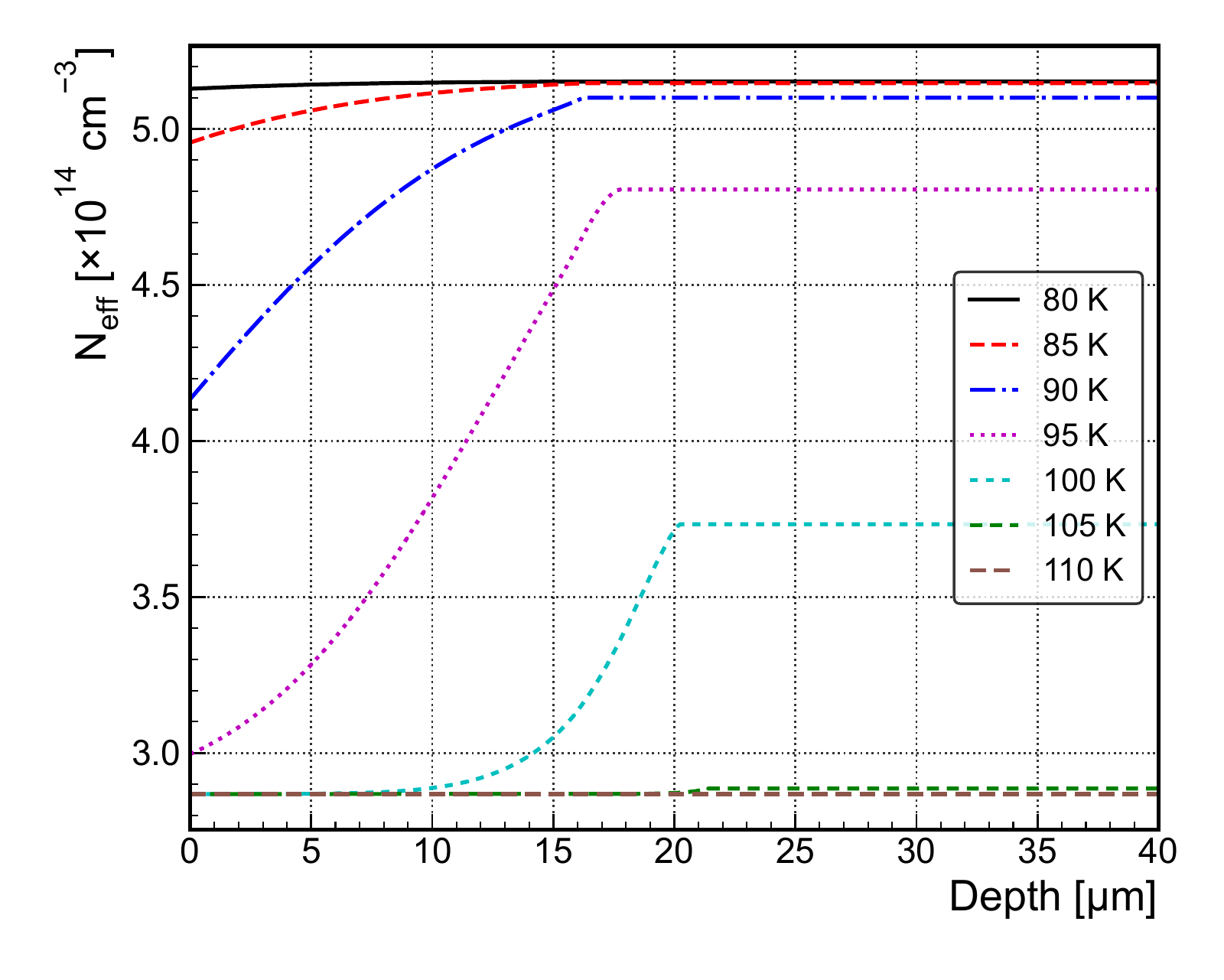}
      } 
\caption{Simulation results of the B$_\text{i}$O$_\text{i}$ generating signals in EPI-7 diode: (a) TS-Cap, comparison with experiment; (b) density of TSC signal, comparison with the measured spectra; (c) and (d) the $E(T,x)$ electric field distribution and the $N_\text{eff}(T,x)$ profiles, respectively, for different temperatures, from 80~K to~110 K, in steps of 5~K. All simulations and given experimental data correspond to a reverse bias of 100~V applied during TS-Cap and TSC temperature scans.}
\label{fig9}
\end{figure*}

\begin{figure*}[htp]
 \centering
  \subfloat[]{
   \includegraphics[width=0.49\linewidth]{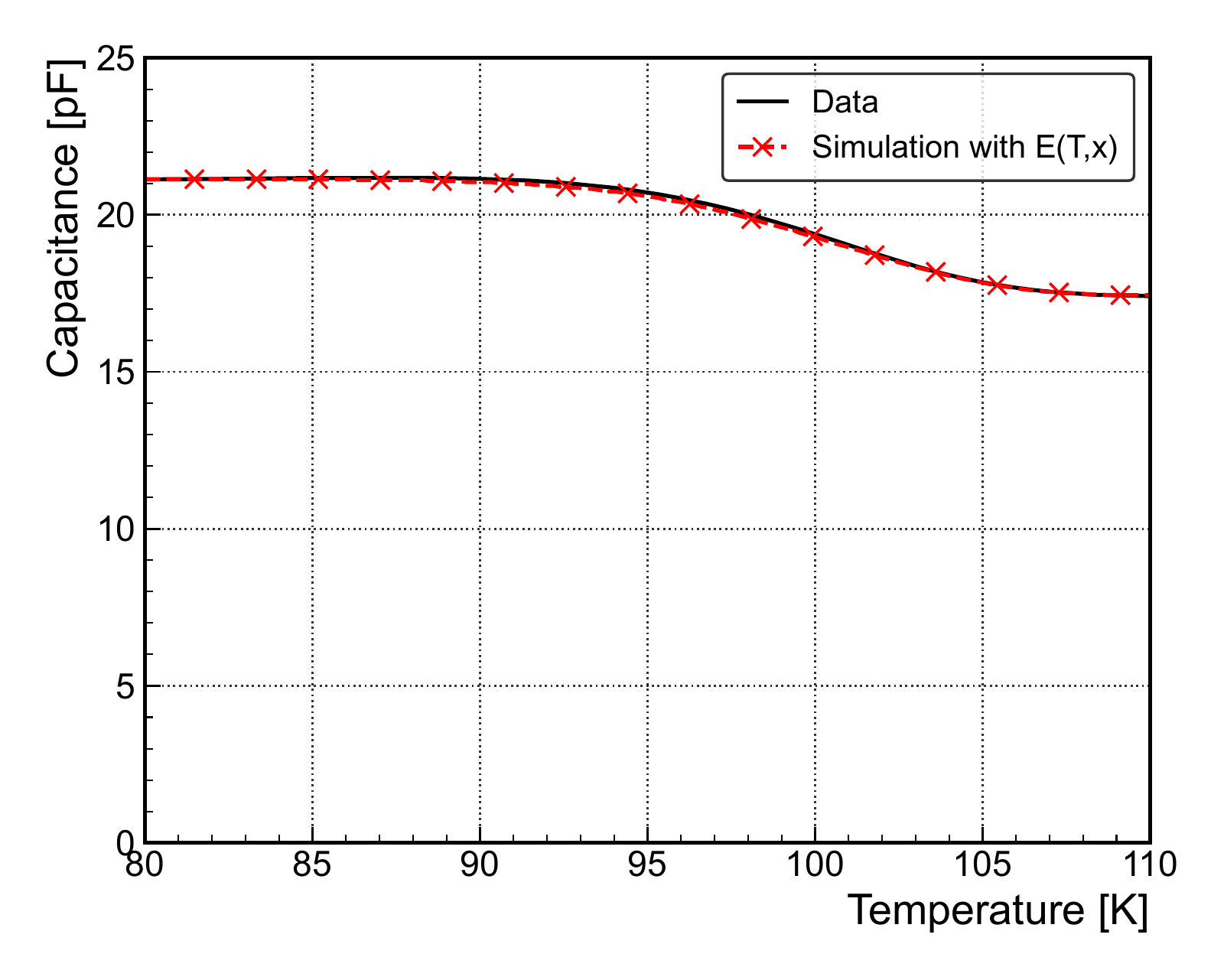}
   }
   \subfloat[]{
   \includegraphics[width=0.49\linewidth]{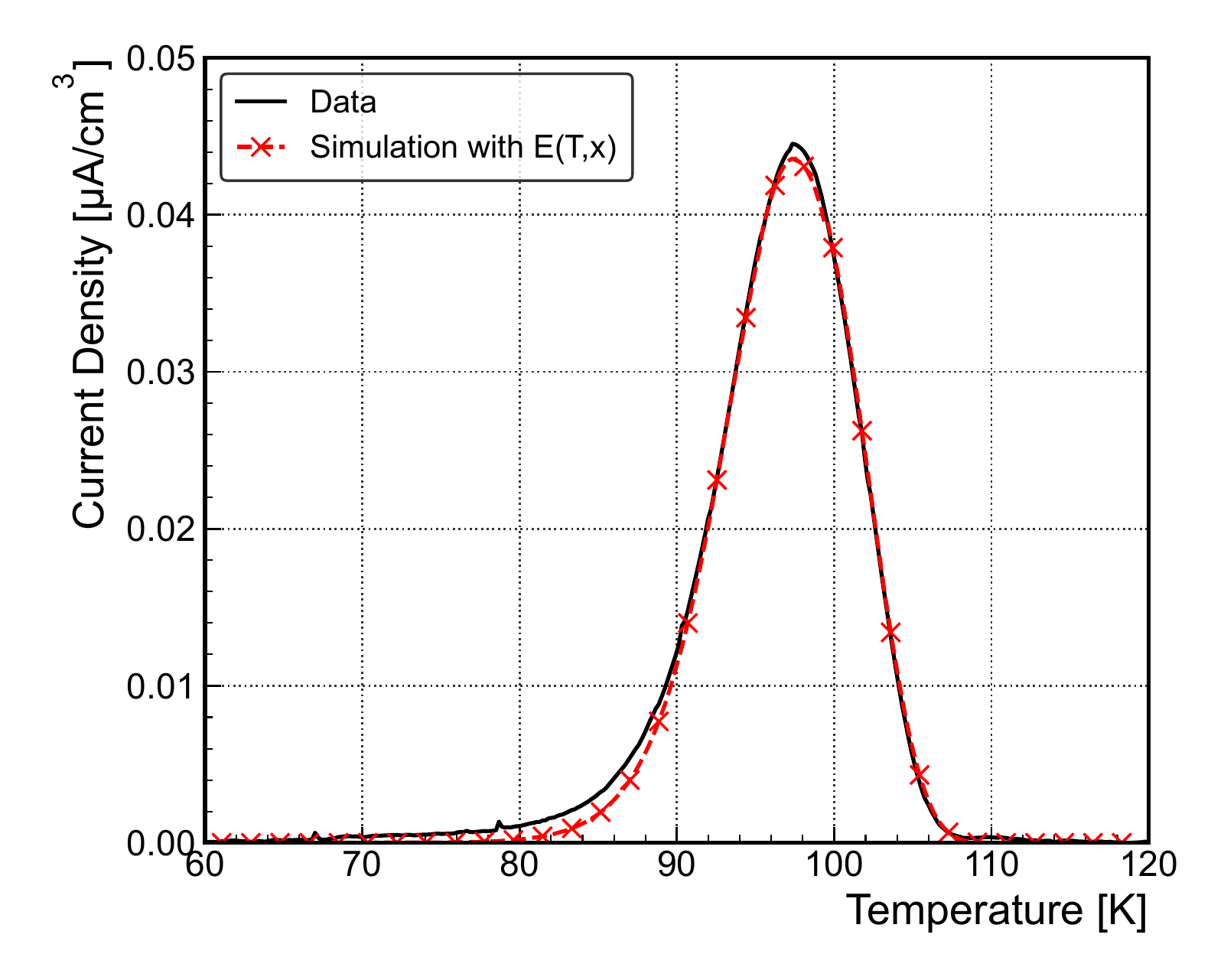}
   } \\
   \subfloat[]{ 
    \includegraphics[width=0.49\linewidth]{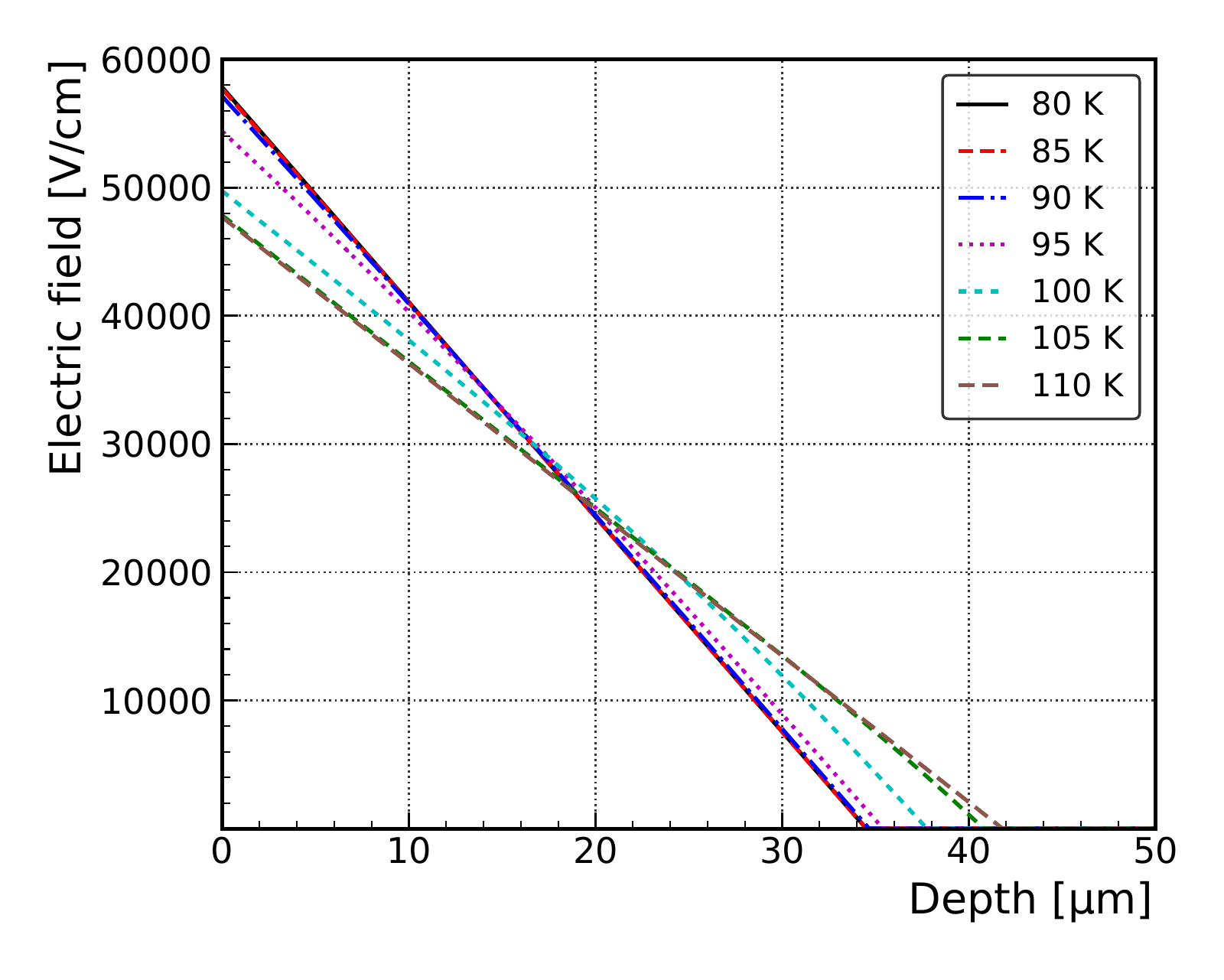}
    }
   \subfloat[]{
      \includegraphics[width=0.49\linewidth]{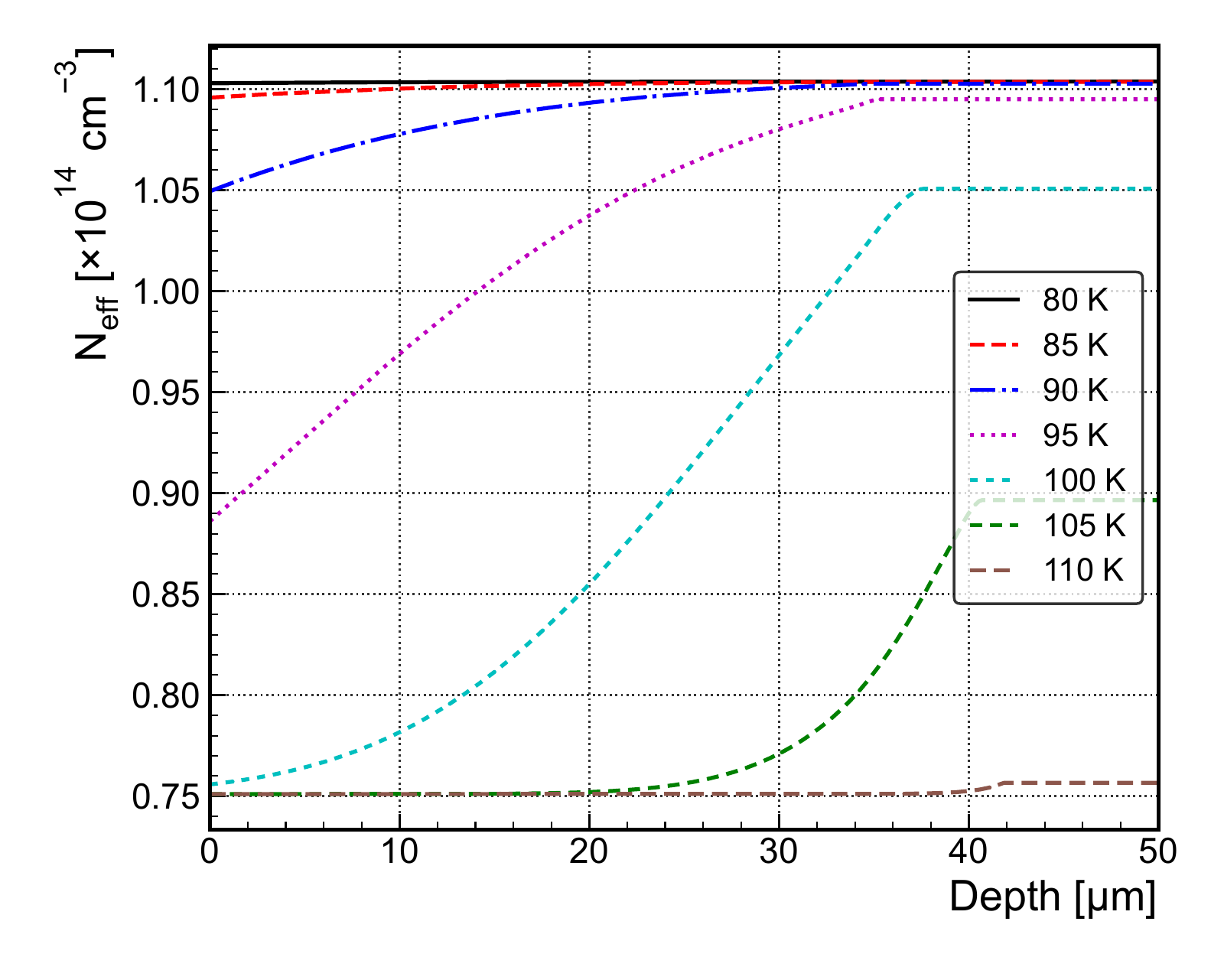}
      } 
\caption{Simulation results of the B$_\text{i}$O$_\text{i}$ generating signals in a 50~$\Omega\cdot$cm EPI diode irradiated with 23 GeV protons to $\Phi_\text{eq}$~=~$4.3 \times 10^{13}$~$\si{cm^{-2}}$: (a) TS-Cap, comparison with experiment; (b) density of TSC signal, comparison with the measured spectra; (c) and (d) the $E(T,x)$ electric field distribution and the $N_\text{eff}(T,x)$ profiles, respectively, for different temperatures, from 80~K to 110~K, in steps of 5~K. All simulations and given experimental data correspond to a reverse bias of 100~V applied during TS-Cap and TSC temperature scans.}
\label{fig10}
\end{figure*}

In the following part, the simulation results and comparison with the corresponding TS-Cap and TSC measurements will be presented for two devices, both annealed for 2~h at 80~$^o$C after irradiation: the electron irradiated sample EPI-7 (see Table~\ref{tab:table1}) and a 50 $\si{\Omega\cdot}$cm $p$-type diode irradiated with 23 GeV protons to $\Phi_\text{eq}$~=~$4.3 \times 10^{13}$~$\si{cm^{-2}}$ for which more detailed information can be found in reference~\cite{b27}. The measurement parameters for both diodes
are the same, i.e. $V_{bias}$  = -100 V, heating rate $\beta$ = 0.183 K/s and the frequency for the capacitance measurement $f$ = 10 kHz. 
\par
All parameters, the fixed and the adjusted ones, used for the simulations of both diodes are summarized in Table~\ref{tab:table2}. For the presented data, the details about $N_{0}$ can be found in the Appendix. The simulation results for the EPI-7 diode are displayed in Fig.~\ref{fig9}~(a-d). In order to reproduce the TS-Cap measurement (Fig.~\ref{fig9}(a)) the B$_\text{i}$O$_\text{i}$ concentration was extracted via the Eq. (14), the $\xi$ value for the Poole-Frenkel effect was set to $\xi$ = 0.5 and the zero-field activation energy $E_{a0}$ = 0.258~eV. With the same values for $\xi$ and $E_{a0}$ parameters but a lower B$_\text{i}$O$_\text{i}$ concentration, the TSC signal could be reproduced in the temperature range between 90 K and 105 K. The low temperature tail, which can not be described by the simulation, is most probably due to the so-called X-defect (see Fig.~\ref{fig5}~(a, b)). Contrary to the TSC case, where the charge emission from the X defect can be separated from that of the B$_\text{i}$O$_\text{i}$ defect, in TS-Cap measurements the contributions of both defects cannot be separated. Therefore, the concentration extracted from the TS-Cap curve is larger compared to the value derived from the TSC spectrum.
\par
Included in Fig.~\ref{fig9}~(a, b) are also the results from simulations which use the position independent average electric field $<E(T)>$ = $V_{bias}/w(T)$ where $w(T)$ is given by: 
\begin{eqnarray} 
w(T)  & = & \sqrt{{{2 \varepsilon_0 \varepsilon_r V_{bias}} \over {q_0 N^{'}_\text{eff}(T)}}}
\end{eqnarray}
Here $N^{'}_\text{eff}(T)$ is constant over the depth of the diode and given by Eq. (13) where $f(T)$ is calculated with the average electric field $<E(T)>$ of the previous temperature step. For this case, the value $\xi$ = 1 and a higher zero-field activation energy of $E_{a0}$ = 0.284~eV is needed in the simulation, in order to get the best fit to the experimental data.
\par

In Fig.~\ref{fig9} (c, d) the electric field distribution and the $N_\text{eff}$ profiles as a function of the depleted depth are plotted for temperatures between 80 K and 110 K in steps of 5 K. As it can be seen in Fig.~\ref{fig9}(c), with increasing the temperature, the maximal value of $E(T_{k}, x=0)$ decreases and the depleted depth increases. This corresponds to the development of the effective space charge density $N^{'}_\text{eff}(T_{k}, x)$ for the different temperature steps as shown in Fig.~\ref{fig9}(d). Further, the distribution of the electric field shows a constant gradient before and after the B$_\text{i}$O$_\text{i}$ emission (below 85~K and above 100~K) and position dependent gradients during emission of the B$_\text{i}$O$_\text{i}$ in the range between 85~K and 100~K. This is due to the non-uniform distributed space charge density resulting from the field dependent emission from the defect energy level.
\par
Similar simulations have been performed for the 23~GeV proton irradiated diode and the results are presented in Fig.~\ref{fig10} (a-d). As it can be seen, the simulation of the TS-Cap signal, shown in Fig.~\ref{fig10}(a), is in excellent agreement with the measured data. In this case, the parameters from the measured $C(T)$ curve, by using the same procedure as for the electron irradiated diode, are $\xi$ = 0.85, $E_{a0}$ = 0.265 eV and [B$_\text{i}$O$_\text{i}$] = $\num{3.5e13}$~$\si{cm^{-3}}$. In Fig.~\ref{fig10}(b) the corresponding TSC data and simulated spectra are given. Also in this case the simulation reproduces the data very well, but compared with the TS-Cap simulations, the best agreement is found for slightly different $E_{a0}$ and [B$_\text{i}$O$_\text{i}$] values, of $E_{a0}$~=~0.273~eV and [B$_\text{i}$O$_\text{i}$]~=~$\num{3.3e13}$~$\si{cm^{-3}}$. The $\xi$ value is the same for both simulations. The distributions of the electric field and the N$_\text{eff}$ profiles are plotted in Fig.~\ref{fig10}(c) and (d) for temperatures between 80~K and 110~K in steps of 5~K. In this case, the maximal electric field $E(T_{k}, x=0)$ also decreases with increasing the temperature while the depleted region depth increases. The main difference to the electron irradiated device is the lower field strength in the bulk.
\par
For getting a better fit to the data, the [B$_\text{i}$O$_\text{i}$] for the simulation of the TSC spectrum (see Fig.~\ref{fig9}(b) and Fig.~\ref{fig10}(b)) is adjusted. The [B$_\text{i}$O$_\text{i}$] extracted from the TSC spectrum by integration from 80~K to 105~K is about 13~\% larger compared to the value used for the simulation. This difference is due to the low temperature tail in the spectrum which was not reproduced in the simulation. For the EPI-7 diode, the significant difference of [B$_\text{i}$O$_\text{i}$] between TS-Cap and TSC is caused by some unknown effect.
\par
In principle, the $E_{a0}$ is known with value in between 0.27-0.28~eV~\cite{b27} with fixed $\sigma_n$ = $\num{1.05e-14}$~$\si{cm^2}$. The difference of $E_{a0}$ between TSC and TS-Cap measurements for proton irradiated diode is due to the difference in the temperature of the peak maximum $T_\text{max}$ in the TSC spectrum and the temperature of the turning point in the TS-Cap curve. The related effect is still unknown. The difference of $E_{a0}$ between the proton and electron irradiated devices might be caused by the different production technology of both devices, the diode with a guard ring (p irradiated) and the other one without. The explanation can be proved by comparing the results from the EPI-diode irradiated with protons (\cite{b27}). This diode has roughly the same $E_{a0}$ as the one presented in Table~\ref{tab:table2} for proton irradiation.  
\par
For EPI-7 diode the difference in $E_{a0}$ between the two different electric field distributions (linear electric field $E(x)$ and homogeneous electric field distribution $<E(T)>$) is 26~meV ($\sim$10\%). The reason for this difference can be understood by the fact that the emission rate $e^{pf}_{n} (T, E(x), E_{a0})$ depends exponentially on the electric field distribution. At a specific temperature $T$, the emission rate is enhancing with $E(x)$ and is decreasing when increasing the $E_{a0}$ values. Thus, for the same bias voltage, the values of $E(T,x)$ in the case of linear field distribution and of the average electric field $<E(T)>$ coincide only in the middle of the depleted width, in the front region of the junction $E(T,x)$ being larger than $<E(T)>$ and in the back side smaller. Consequently, the same measured TSC signal can be reproduced in both cases if in the calculation of the emission rates the values of $E_{a0}$ and $\xi$ are smaller for a linear distribution of the electric field than for the constant, average one. This has with respect to the emission rate to be compensated by a lower $E_{a0}$ or a lower $\xi$ value compared to the constant field case in order to reproduce the same measured TSC signal, as it can be seen in Table~\ref{tab:table2}.
\par
Due to the fact that by using Eq.~(16) the experimental data could not be reproduced, a constant $\xi$ was introduced for modifying the field dependence in the Poole-Frenkel effect. The $\xi$ values are different for linear and constant electric fields, 0.5 and 1.0, respectively, while for each $E(x)$ distribution they are the same for simulating the TSC and TS-Cap data. 
\par

\section{Conclusion}
In this work investigations of radiation damage of silicon diodes manufactured on p-type EPI- and Cz-material with a resistivity of about 10~$\Omega\cdot$cm and exposed to \SI{5.5}{\mega\electronvolt} electrons of different fluence values ($1 \times 10^{15}$, $4 \times 10^{15}$, $6 \times 10^{15}$~$\si{cm^{-2}}$) have been performed. The macroscopic properties of the devices, the leakage current density $J_d$ and $N_\text{eff}$, were obtained from I-V and C-V measurements. The microscopic properties of the B$_\text{i}$O$_\text{i}$ and C$_\text{i}$O$_\text{i}$ defects were studied using the TSC and TS-Cap methods and the results are discussed in connection with Boron removal process observed in macroscopic measurements.
\par
The main results obtained in this study are:
\begin{enumerate}
 \item[a)] The density of leakage current $J_d$ increases linearly with the achieved fluence and the corresponding current related damage parameter is determined to be $\alpha$~=~$3.2 \times 10^{-19}$~A/cm. Such a small value was also reported for n-type silicon diodes after irradiation with \SI{5.5}{\mega\electronvolt} electrons~\cite{b22}. Compared with hadron irradiation, the obtained $\alpha$ parameter is much smaller, indicating that the increase of the leakage current caused by low energy electrons is substantially less than that caused by hadrons. Also, the change of $J_d$ with annealing time at 80~$^o$C is strongly suppressed compared with hadron irradiated devices indicating that the irradiation with low energy electrons creates less current generation centers and more stable defects.
 \item[b)] The $N_\text{eff}$ decreases nearly linear with increasing fluence and remains stable during the isothermal annealing at 80~$^\text{o}$C, in agreement with the thermal stability of the B$_\text{i}$O$_\text{i}$ defect~\cite{b24}.
 \item[c)] The development of B$_\text{i}$O$_\text{i}$ and C$_\text{i}$O$_\text{i}$ defects with fluence is linear, however, with different introduction rates for EPI and Cz materials, due to the different Carbon content in the two materials (more in Cz than in EPI) and the competing reactions between Boron and Carbon interstitials with abundant Oxygen interstitials in silicon. Thus, while the introduction rate of B$_\text{i}$O$_\text{i}$ is much smaller in Cz than in EPI material, of 0.63~$\si{cm^{-1}}$ compared with 1.75~$\si{cm^{-1}}$ as seen in Fig.~\ref{fig7}(b), the opposite is happening for C$_\text{i}$O$_\text{i}$. Similar behaviour was also reported in the RD50 collaboration program~\cite{b9,b35}.
 \item[d)] The formation of B$_\text{i}$O$_\text{i}$ defect is the main cause for the change seen in $N_\text{eff}$ after irradiation with \SI{5.5}{\mega\electronvolt} electrons. This was nicely evidenced in EPI diodes where the homogeneous Boron doping profile allowed accurate evaluations. Thus, by comparing the Boron removal rate of 3.7~$\si{cm^{-1}}$ resulted from C-V measurements with that of 3.5~$\si{cm^{-1}}$ resulted by accounting twice the value of B$_\text{i}$O$_\text{i}$ introduction rate due to the donor character of the defect, a good agreement is obtained.
 \item[e)] The TS-Cap technique proved to be a valuable complementary to the TSC tool in order to accurately characterize the radiation induced defects in highly irradiated and partially depleted silicon sensors. This is especially important in the case of low resistivity diodes when the total depletion of the device in TSC measurements cannot be achieved or the depletion depth cannot be kept constant during the temperature scan. However, TS-Cap allows the evaluation of defect concentrations only if the defects are well isolated in the silicon bandgap, not overlap with other defects. 
 \item[f)] The temperature dependence of the thermally stimulated capacitance at constant bias voltage and of the corresponding TSC spectra, for a \SI{5.5}{\mega\electronvolt} electron and a 23~GeV proton irradiated devices were simulated in the temperature range of the B$_\text{i}$O$_\text{i}$ defect emission. For reproducing the TS-Cap and TSC data the Poole-Frenkel effect was accounted and modified by a subunitar factor $\xi$ and small variations in the defect's zero-field activation energy. Different $\xi$ and $E_\text{a0}$ values resulted from simulating the experimental data measured on differently damaged silicon diodes, an aspect that has to be further studied in more detail. Presently, we justify these adjustments by the fact that the Poole-Frenkel theory was not developed for position dependent electric fields as existing in diodes and more pronounced in low resistivity ones, but for constant field around the defect. In the absence of a proper Poole-Frenkel theory for accounting the position dependent electric field in diodes, the adjustments were made for describing as good as possible both B$_\text{i}$O$_\text{i}$ current and capacitance signals. In addition, when accounting for the electric field dependent electron emission from B$_\text{i}$O$_\text{i}$ defect, the simulated electric field distributions $E(T, x)$ for the temperature range where the B$_\text{i}$O$_\text{i}$ defect discharges, between 80~K and 110~K, show position dependent gradients, corresponding to the position dependent effective space charge densities $N^{'}_\text{eff}(T, x)$.
\end{enumerate}

\appendix
\section{Simulation}
\begin{figure}[!htb]
  \centering
    \subfloat[]{
     \includegraphics[width=1\linewidth]{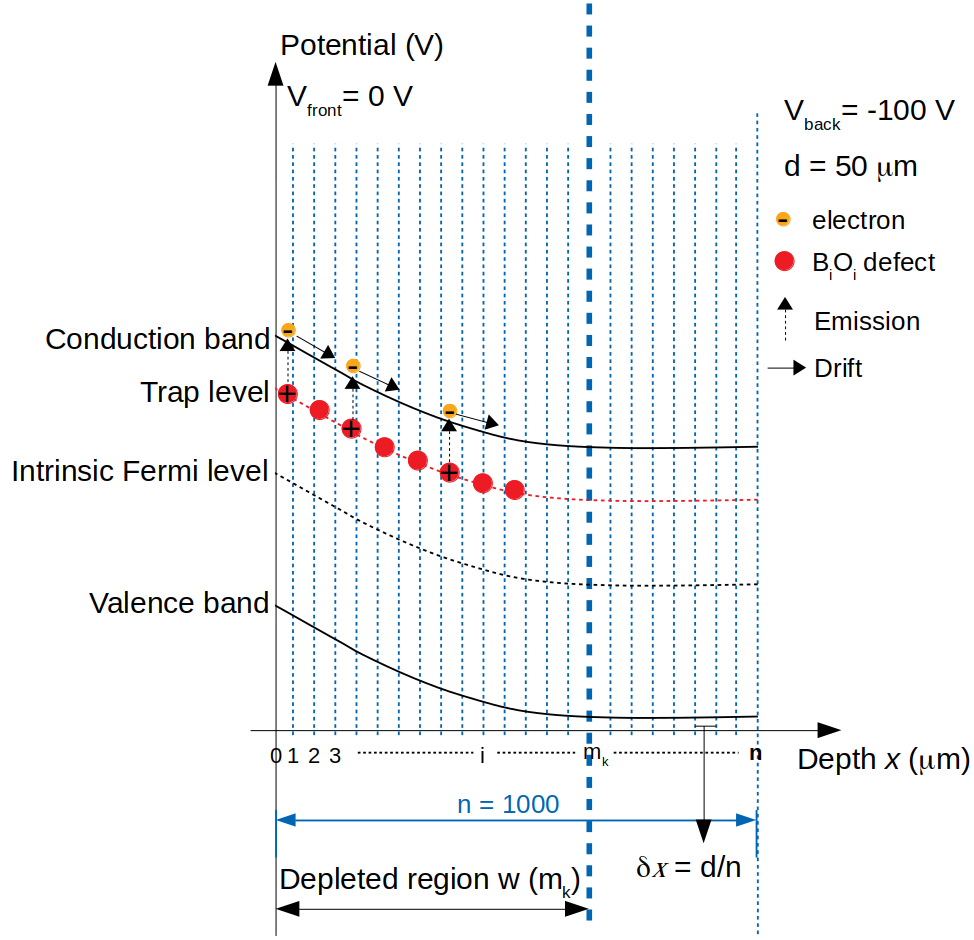}     
    }\\
    \subfloat[]{
     \includegraphics[width=1\linewidth]{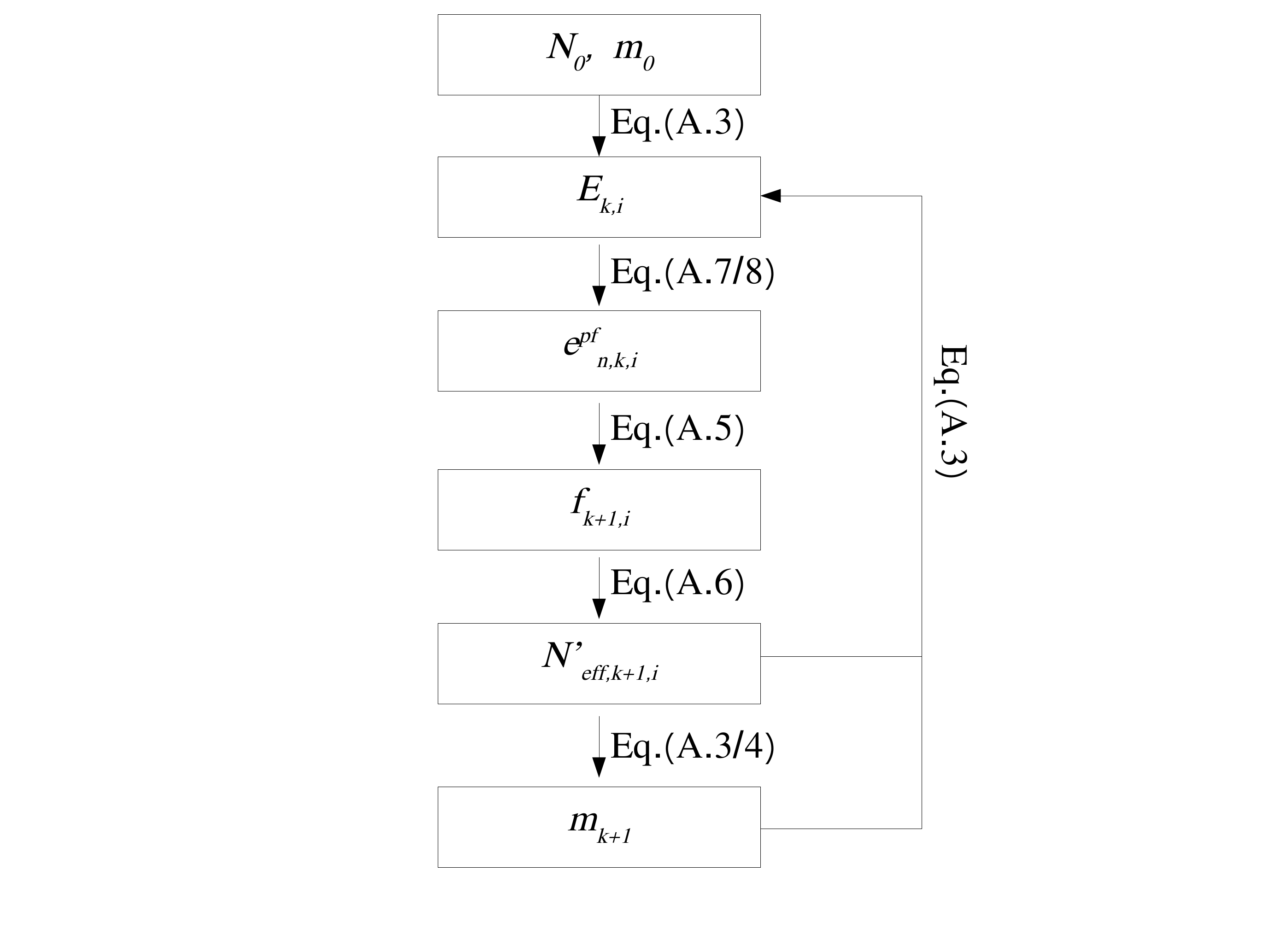}
    }
\caption{(a) Schematic of finite elements approach applied for describing the field dependent B$_\text{i}$O$_\text{i}$ emission in TS measurements. The band gap structure has been divided into n layers by blue lines. (b) simulation procedure.}
\label{fig11}
\end{figure}

In this section, the simulation procedure for the TS-Cap and the TSC spectra of the B$_\text{i}$O$_\text{i}$ defect will be described. The simulations are performed by using Python software. The bulk of the sensor is divided into n sufficiently thin layers of a thickness $\delta x$ = d/n, where $d$ is the thickness of the EPI- or Cz-silicon (see Fig.~\ref{fig11}a)). The index $i$ in Fig.~\ref{fig11}a) runs from 0 to n and the boundary between the depleted and
the non-depleted region is labelled $m_k$. The index $k$ indicates the
temperature step $T_k$ and varies between 0 (the start temperature $T_0$) and
$f$ (the final Temperature $T_f$). 
\par
As the emission rate of the B$_\text{i}$O$_\text{i}$ defect is governed by the 3-D Poole-Frenkel effect (Eq. (15, 16, 17)) the electric field distribution $E(T,x)$ has to be calculated via the  Poisson equation (Eq. (19)) for a known effective space charge density $N'_\text{eff}(T,x)$ (Eq. (18)). Considering the finite element method mentioned above, the Poisson equation can be written as:
\par
\begin{flalign}
E_{k,i+1} - E_{k,i} = \frac{q_0 N^{'}_{\text{eff},k,i}}{\varepsilon_0 \varepsilon_r}
\cdot \frac{d}{n}
\end{flalign}
Considering the boundary condition between the depleted and the non-depleted region
\par

\begin{equation} 
E_{k,i} = 0~~~\text{for}~~~i \geq m_k ,
\end{equation}
the $E_{k,i}$ is given by (according to Eq.~(A.1)):
\par
\begin{equation} 
E_{k,i} = \sum_{j=0}^{m_k} {{q_0 N^{'}_{\text{eff},k,j}} \over {\varepsilon_0
\varepsilon_r}} \cdot {d \over n} - \sum_{j=0}^{i} {{q_0 N^{'}_{\text{eff},k,j}} \over
{\varepsilon_0 \varepsilon_r}} \cdot {d \over n},
\end{equation}
where the index $j$ is used to sum $N{'}_{\text{eff},k,i}$ for layer $i$ from 0 to the boundary or the indicated layer $i$. Further, the applied bias voltage $V_{bias}$ is given by the sum of all electric field steps $E_{k,i}$ up to the temperature dependent $m_k$ value, as given
by:
\par
\begin{equation}
V_{bias} = \sum_{i=0}^{m_k} E_{k,i} \cdot \frac{d}{n} 
\end{equation}
This equation (Eq. A.4)) is then used to calculate $m_k$ by rising $m_k$ from 0 up to the value that fulfils Eq. (A.4). Thus for the description of Eq.(A.1-4), the only unknown parameter for obtaining the $E_{k,i}$ is $N^{'}_{\text{eff}, k,i}$. In Eq.(18), $N^{'}_{\text{eff}, k, i}$ can be obtained by $f_{k, i}$, which in finite elements method can be written as:
\par
\begin{flalign}
f_{k+1,i} = exp\left(-\sum_{j=0}^{k} {{\Delta T_j \over {\beta} }e_{n,j,i}}\right),
\end{flalign}
where the index $j$ was used to sum the emission rate from temperature $T_0$ to the $T_k$. The Eq.~(18) can be changed to:
\par
\begin{flalign}
N^{'}_{\text{eff},k,i} = N_{0} - [\text{B}_\text{i}\text{O}_\text{i}] \cdot (1-f_{k,i})
\end{flalign}
Considering the 3-D Poole Frenkel effect, the emission rate can be written to:
\par
\begin{flalign}
e^{pf}_{n,k,i} = \sigma_n v_{th,n} N_c exp\left(-\frac{E_{a0}}{k_B T_k}\right) 
\left[\left(\frac{1}{\gamma^2_{k,i}}\right) \left(e^{\gamma_{k,i}}
\left(\gamma_{k,i}+1\right)\right)+ \frac{1}{2}\right]
\end{flalign}
with
\par
\begin{equation} 
\gamma_{k,i} = \xi \cdot \sqrt{q_0 \lvert {E_{k,i}} \rvert \over {\pi \varepsilon_0
\varepsilon_r}} \cdot {q_0 \over {k_B T_k}} 
\end{equation}
\par
For the electron capture cross section we used the value of $\sigma_n$~=~$\num{1.05e-14}$~$\si{cm^2}$ determined experimentally in~\cite{b33}. The zero field activation energy of the B$_\text{i}$O$_\text{i}$ defect, $E_{a0}$, was previously determined to be between 0.271~eV and 0.288~eV for silicon diodes with resistivities varying from 50~$\Omega\cdot$cm to 2~k$\Omega\cdot$cm and irradiated with 23~GeV protons~\cite{b27}. $E_{a0}$ values were tuned for getting the best fit between simulated and measured data, and all parameters used in the simulation are given in Table~\ref{tab:table2}. The concentration [B$_\text{i}$O$_\text{i}$] used for TS-Cap simulation was extracted
from the TS-Cap measurement according to Eq.~(14). 
\par
The initial conditions for $T_k$, $N^{'}_{\text{eff},k,i}$, $e^{pf}_{n,k,i}$, $f_{k,i}$, $n$ and the applied bias voltage $V_{bias}$ are: $T_0$ = 40 K, $N^{'}_{\text{eff},0,i}$ ($N_0$) as extracted from TS-Cap at 80~K, $e^{pf}_{n,0,i}$ = 0, $f_{0,i}$ = 1, $n$ = 1000 and $V_{bias}$ = -100 V. Then, it is obtained that for initial $T_0$ electric field distribution $E_{0,i}$ decreases linearly from $\num{1.2e5}$ and $\num{5.8e4}$ to 0 $\si{V/cm}$ for 5.5~MeV electrons with $\Phi_\text{e} = \num{4e15}$~$\si{cm^{-2}}$ and 23 GeV protons $\Phi_\text{p} = \num{6.91e13}$~$\si{cm^{-2}}$ irradiation, respectively. It was also given that the $m_0$ is equal to 320 and 690 for electrons and protons irradiation, respectively. Such values were extracted from Eq.~(A.3) and Eq.~(A.4). Next, these values were used to calculate the emission rate $e^{pf}_{n,1,i}$, $f_{1,i}$ and $N^{'}_{\text{eff},1,i}$ (Eq.~(A.5-8)), with which the distribution of the electric field $E_{1,i}$ and $m_1$ are calculated. This step by step calculation continues until the final temperature $T_f$ is reached. Also, the temperature dependent depletion depth is calculated according to $w(T_k)$ = $m_k \cdot d/n$. The selected $T_e$ must be higher than the temperature of the end of emission, and in this work $T_e$ = 120~K was chosen for simulation. The TSC values at $T_k$ can also be calculated according to:
\begin{eqnarray} 
I^{e}_{TSC, k}  & = & q_0 A [B_i O_i] \sum_{j=0}^{m_k} {{j \cdot {d \over n}} \over {w(T_k)}} \cdot e^{pf}_{n,k,j}  \cdot  f_{k,j} \cdot {d \over n}
\end{eqnarray}
Considering the depleted depth extracted from the TS-Cap measurements, the TSC spectrum was also simulated and compared with the measured data.

\section*{Acknowledgment}
This work has been carried out in the framework of the RD50 Collaboration.
The project has received funding from the European Unions Horizon 2020 Research and Innovation program under Grant Agreement no.\ 654168. C.~Liao would like to thank for the given support to work at the Hamburg University to the Deutsche Forschungsgemeinschaft (DFG, German Research Foundation) under Germany's Excellence Strategy -- EXC2121 "Quantum Universe" -- 390833306 project and to Professor Z.~Li. I.~Pintilie and Lucian~D.~Filip acknowledge the funding received through IFA-CERN-RO 08/2022 project and Core Program 2019-2022 (contract 21N/2019). Z.~Li acknowledges the funding received through the Key Scientific and Technological Innovation Project of Shandong Province under Grant No. 2019 TSLH 0316, and the project of Yantai Institute for the exchange of Driving Forces under Grants No. 2019XJDN002.

\bibliographystyle{elsarticle-num-names}

\end{document}